\documentclass[12pt,letterpaper]{JHEP3}
\usepackage{amsmath,epsfig}

\newcommand{\st}{{\tilde{t}}}
\newcommand{\sta}{{\tilde{t}_1}}

\newcommand{\mst}{m_{\tilde{t}_1}}
\newcommand{\Dmst}{\Delta m_{\tilde{t}_1}}

\newcommand{\neu}{\tilde{\chi}^0}
\newcommand{\mcha}[1]{m_{\tilde{\chi}^\pm_{#1}}}
\newcommand{\mneu}[1]{m_{\tilde{\chi}^0_{#1}}}

\newcommand{\stopone} {\tilde{t}_1}
\newcommand{\stoponestar} {\tilde{t}_1^*}
\newcommand{\Nth} {N_{\mathrm{th}}}
\newcommand{\Bth} {B_{\mathrm{th}}}
\newcommand{\Npk} {N_{\mathrm{pk}}}
\newcommand{\Bpk} {B_{\mathrm{pk}}}
\newcommand{\rtsth} {\sqrt{s_{\mathrm{th}}}}
\newcommand{\rtspk} {\sqrt{s_{\mathrm{pk}}}}
\newcommand{\Lumth} {{\cal{L}}_{\mathrm{th}}}
\newcommand{\effpk} {\epsilon_{\mathrm{pk}}}
\newcommand{\effth} {\epsilon_{\mathrm{th}}}
\newcommand{\Lumpk} {{\cal{L}}_{\mathrm{pk}}}

\newcommand{\Lum}   {{\cal{L}}}
\newcommand{\eff}   {\epsilon}
\newcommand{\Nev}   {N_{\mathrm{ev}}}
\newcommand{\pbinv} {{\mathrm{pb}^{-1}}}
\newcommand{\fbinv} {{\mathrm{fb}^{-1}}}
\newcommand{\fb}    {{\mathrm{fb}}}
\newcommand{\efftrue}   {\epsilon^{\mathrm{true}}}
\newcommand{\effest}    {\epsilon^{\mathrm{est}}}
\newcommand{\Ytrue}     {Y^{\mathrm{true}}}
\newcommand{\Yest}      {Y^{\mathrm{est}}}
\newcommand{\deltath}   {\delta_{\mathrm{th}}}
\newcommand{\deltapk}   {\delta_{\mathrm{pk}}}
\newcommand{\Evis}      {E_{\mathrm{vis}}}
\newcommand{\Mvis}      {M_{\mathrm{vis}}}
\newcommand{\pt}        {p_T}
\newcommand{\Ntracks}   {N_{\mathrm{tracks}}}
\newcommand{\Njets}     {N_{\mathrm{jets}}}
\newcommand{\costhetathrust} {\cos\theta_{\mathrm{thrust}}}

\newcommand{\Mjj}            {m_{\mathrm{jj}}}
\newcommand{\Mjjsq}          {m_{\mathrm{jj}}^2}
\newcommand{\Pcharm}         {P_{\mathrm{c}}}
\newcommand{\qqbar}          {q\bar{q}}
\newcommand{\cflag}          {F^{(c)}_i}
\newcommand{\cflagone}       {F^{(c)}_1}
\newcommand{\cflagtwo}       {F^{(c)}_2}
\newcommand{\epsilonstop}    {\epsilon_{\tilde{t}}}
\newcommand{\etal}           {{\it et~al.}}

\def\mathswitch#1{\relax\ifmmode#1\else$#1$\fi}
\def\mathswitchr#1{\relax\ifmmode{\mathrm{#1}}\else$\mathrm{#1}$\fi}

\newcommand{\gev}{\,\, \mathrm{GeV}}
\newcommand{\gevsq}{\,\, \mathrm{GeV}^2}

\newcommand{\lesim}{\,\raisebox{-.1ex}{$_{\textstyle <}\atop^{\textstyle\sim}$}\,}

\newcommand{\epem}{e^+e^-}
\newcommand{\mpmm}{\mu^+\mu^-}

\title{A Method for the Precision Mass Measurement of the Stop Quark
at the International Linear Collider}

\author{Ayres Freitas\\
University of Chicago, Chicago, IL 60637, USA \\
HEP Division, Argonne National Laboratory, Argonne, IL 60439, USA   \\
Institut f\"ur Theoretische Physik,
        Universit\"at Z\"urich, 8057 Z\"urich, Switzerland\\
E-mail: \email{afreitas@physik.unizh.ch}}
\author{Caroline Milst\'ene\\
Fermi National Accelerator Laboratory, Batavia, IL 60510-500, USA\\
E-mail: \email{caroline@fnal.gov}}
\author{Michael Schmitt\\
Northwestern University, Evanston, IL 60208, USA\\
E-mail: \email{schmittm@lotus.phys.northwestern.edu}}
\author{Andre~Sopczak\\
Lancaster University, Lancaster LA1 4YB, United Kingdom\\
E-mail: \email{andre.sopczak@cern.ch}}

\abstract{
Many supersymmetric models predict new particles within the reach of the
next generation of colliders. For an understanding of the model structure and
the mechanism(s) of symmetry breaking, it is important to know the masses of the
new particles precisely. In this article the measurement of the mass of
the scalar partner of the top quark (stop) at an $e^+e^-$ collider is studied.
A relatively light stop is motivated by attempts to explain electroweak
baryogenesis and can play an important role in dark matter relic density.
A method is presented which makes use of cross-section measurements near the
pair-production threshold as well as at higher center-of-mass energies. It is
shown that this method not only increases the statistical precision, but
also greatly reduces the systematic uncertainties, which can be important.
Numerical results are presented, based on a realistic event simulation, for two
signal selection strategies: using conventional selection cuts, and using an
Iterative Discriminant Analysis (IDA). Our studies indicate that a precision
of $\Delta\mst = 0.42$~GeV can be achieved, representing a major improvement
over previous studies.  While the analysis of stops is particularly challenging 
due to the possibility of stop hadronization, the general procedure could be applied 
to the mass measurement of other particles as well.
We also comment on the potential of the IDA to discover a stop quark
in this scenario, and we revisit the accuracy of the theoretical predictions
for the neutralino relic density.
}


\preprint{
ANL--HEP--PR--07--107
\\
FERMILAB--PUB--07--185--E
\\
NUHEP--EXP/07-02
\\
ZH--TH 15/07
}

\begin{document}

\section{Introduction}


An experiment at the International Linear Collider (ILC) 
will be able to make many precise 
measurements from which particle properties, and ultimately,
the outlines of a particle physics model may be inferred.
Due to the high statistical precision expected at the ILC, the optimization 
of the systematic errors is of particular importance.
We have studied one specific example, namely, the extraction
of the mass of an hypothetical stop squark from cross-section
measurements near threshold.  We have devised a method which
reduces most systematic uncertainties and leads to a potentially
very accurate measurement of the stop squark mass.  This method,
however, is general and could be applied to other particles
produced in an $\epem$ collider.

The method relies on the comparison of production rates at two different
center-of-mass energies, and knowledge of how the cross-section
varies as a function of $\sqrt{s}$ and the particle mass.
In simple terms, one measures the yield at an energy close to 
the pair-production threshold, which will be very sensitive
to the particle mass, and then at a much higher energy,
which has little sensitivity.  The ratio of these two yields
retains sensitivity to the mass, and at the same time is
{\it insensitive} to many potential systematic effects

We have chosen the case of a light scalar top squark with a mass not much higher
than the mass of the lightest neutralino since production of this particle was
already extensively  studied in an ILC
context~\cite{stop,stopsLC}.   It was concluded
that a conventional approach to the measurement of the stop squark mass
culminated in an uncertainty of about $\Dmst = 1.2\gev$~\cite{stop,heavyq}. 
The new method improves substantially on this result, and for a similar 
scenario, we conclude that the uncertainty will be $\Dmst = 0.42$~GeV. 

For this analysis, we have performed realistic simulations of the 
signal and backgrounds, and used two techniques to separate the 
signal from the background. The first technique is based on conventional 
selection cuts, while the second employs an improved Iterative 
Discriminant Analysis (IDA)~\cite{ida}. Furthermore, the hadronization of the 
stop has been included and we have carefully studied the systematic 
uncertainties arising from this and other sources.

There are theoretical motivations for studying a light 
stop squark with a mass close to the neutralino mass.  
Specifically, we evoke a scenario within the Minimal Supersymmetric
extension of the Standard Model (MSSM) which is able to explain the
dark matter density of the universe as well as the baryon
asymmetry through the mechanism of electroweak baryogenesis.
The existence of dark matter has been firmly established
by various observations, most notably by the measurements
of the cosmic microwave background radiation by the
Wilkinson Microwave Anisotropy Probe~(WMAP)~\cite{wmap} 
and the studies conducted by the Sloan Digital Sky 
Survey~\cite{Tegmark:2003ud}. 
The known properties of dark matter suggest that it consists
of primordial weakly-interacting massive particles.
Within the context of supersymmetry, the best candidate
is the lightest neutralino, $\neu_1$, which is generically
the lightest supersymmetric particle, and would be 
stable if $R$-parity is conserved.

Another well-established fact which poses a great puzzle
for particle physics is the apparent asymmetry
between the amount of matter and anti-matter in the universe.
There are several competing theoretical explanations for
the origin of this baryon asymmetry.  One of these relies
on asymmetries generated during the electroweak
phase transition.  The hypothesized mechanism is not viable
within the Standard Model (SM), but is possible within the
context of supersymmetry.  In fact, requiring that the correct
baryon asymmetry is generated at the electroweak phase 
transition places strong constraints on the parameter
space of the MSSM~\cite{CQW,EWBG,Carena:1997ki,EWBG2}.
In particular, the lightest scalar top squark $\st_1$ must not be heavy,
satisfying the bound $\mst \lesim 140$ GeV with concomitant bounds
on the mass of the Higgs boson~\cite{Carena:1997ki,EWBG2}.
Furthermore, this particle is predominantly of the right-handed chirality
state. A small mass difference between the stop and the lightest
neutralino can help to bring the dark matter relic density into the proper
range due to co-annihilation between the stop and the neutralino. For this
mechanism to be effective, the typical mass difference is rather small,
$\mst - \mneu{1} \lesim 30\gev$~\cite{Balazs:2004bu}. The dominant
decay mode of the stop is $\st_1 \to c\,\neu_1$, resulting in a final
state with two soft charm jets and missing energy.  Previous
studies~\cite{stop,stopsLC} have shown that clean samples of
such events can indeed be isolated at the ILC.

This paper is organized as follows. 
The next section explains the ratio-of-yields method in detail.
Section~\ref{sec:ana} describes the tools and  methods used for 
simulating the relevant processes and the detector, as well as
two methods for selecting a clear stop signal.
Section~\ref{sec:exp} is devoted to a discussion of the experimental
systematics, followed by Section~\ref{sec:th} which explores
theoretical uncertainties.  The last section reports the results
for this specific channel, and shows the implications for 
future calculations of dark matter relic densities based in
particle physics, specifically, supersymmetry.  
We comment briefly on the potential of the IDA method to
discover this stop quark at $\sqrt{s} = 500\gev$.
Conclusions follow.


\section{Method}
\label{sec:method}
\par
One way to measure the stop mass would be through kinematic distributions of
its final state products. However, jet energies are difficult to measure
precisely, especially when the jets are not energetic.  Furthermore, the radiation
of gluons and the hadronization of the stop quarks complicate the kinematics
in ways that are difficult to predict and model accurately.
These effects make a precise stop mass measurement from kinematic distributions
rather difficult~\cite{stopsLC}.
\par
Alternatively, one can extract the stop mass and mixing angle from measurements
of the cross-section. For example, it has been shown that using measurements
with two different beam polarization at one center-of-mass energy, both the
stop mass and mixing angle can be inferred with good accuracy
\cite{stopsLC}.   For light stop quarks with masses
${\cal O}(100 \gev)$, the typical achievable precision
is $\Dmst \sim 1 \gev$. However, this technique is limited by
substantial systematic uncertainties on the measurement of the total
cross-section, in particular the modeling of stop hadronization and
the resulting uncertainties in the selection efficiency.
\par
We propose a new method which reduces the impact of these systematic 
uncertainties, and which we describe in this section in general terms.  
While our explication is based on the case of a light stop, the method
could be applied to other particles.  (See, for example, Ref.~\cite{snumass}
for a discussion of the sensitivity to unknown branching ratios.)
The original presentation of this method concerned Higgs production
at a future $\gamma\gamma$-collider~\cite{gagaHiggs}.
\par
We want to extract the mass ($M_X$) of a particle from measurements of
its production cross sections.  In order to obtain the best result,
two issues must be considered:
\begin{enumerate}
\item
optimization of the energy and luminosity for the minimum statistical error, and
\item
reduction of systematic uncertainties.
\end{enumerate}
The method described here seeks to address both issues in the best possible way.
\par
The error on the extracted mass ($\Delta M_X$) relates to the 
cross-section measurement error ($\Delta\sigma$) through
\begin{equation}
\label{eq:dMX}
  \Delta M_X = 
  \left| {\frac{d\sigma}{d\, M_X}} \right|^{-1}
  \Delta\sigma .
\end{equation}
It is important to keep in mind that the statistical component
in $\Delta\sigma$ depends on $\sigma$.
\par
For particles pair-produced mainly in the $s$-channel, the tree-level 
cross section depends on the mass through the phase space, which usually 
shows up as factors of the velocity of the particle:
$
  \beta = \sqrt{ 1 - (M_X/E_b)^2 }
$
where $E_b = \sqrt{s}/2$ is the beam energy -- hence, the maximum energy
the given particle can have.
For the pair-production of scalar particles, $\sigma \propto \beta^3/s$,
and for fermions, $\sigma \propto \beta/s$.  These simple rules
can be modified by radiative corrections, and by beam energy spread,
but the basic picture does not change dramatically.  We can use this to
frame the discussion of the statistical error.
\par
It is instructive to minimize $\Delta M_X$ as in Eq.~(\ref{eq:dMX})
with a simple Ansatz $\sigma = \sigma_0\beta^3/s$.
We imagine that $M_X$ is already known approximately, and we want
to select the beam energy at which to run the linear collider 
such that $\Delta M_X$ is minimized, for a given integrated
luminosity~$\Lum$ and selection efficiency~$\eff$.
One easily finds that $\Delta M_X = (s^2/12\sigma_0 M\beta) \Delta\sigma$.
Ignoring systematic uncertainties, one might naively expect that
$\Delta\sigma$ is proportional to $\sqrt{\Nev}$, where 
$\Nev = \sigma\eff\Lum$ is the number of selected events, 
which gives us $\Delta M_X \propto \sqrt{\beta}$.
This surprising result indicates that zero uncertainty on the mass is obtained
at the point at which the signal cross-section vanishes.
\par
The fallacy comes in assuming that  $\Delta\sigma$ is proportional to 
$\sqrt{\Nev}$, which certainly does not apply as $\Nev\rightarrow 0$,
even in the absence of background.  The transition from a region in which
the cross-section is being measured ($\Delta\sigma \propto \sqrt{\Nev}$)
to a region in which an upper limit is being set ($\Nev \lesim 3$)
is discussed clearly in Ref.~\cite{FeldmanCousins}.  One must construct
a confidence belt in the $(M_X,\sigma)$ plane, for a given confidence
level ---  68\% would be appropriate for a measurement.  In the present
case, this belt will depend on $\sigma(M_X)$, as well as on~$\eff\Lum$.
When the expected value for $\Nev$ becomes too small, there is no 
upper bound on~$M_X$, and Eq.~(\ref{eq:dMX}) clearly does not apply.
In addition, an account of background estimates, of experimental 
uncertainties on~$\eff$ and~$\Lum$, and theoretical uncertainties
on $\sigma(M_X)$ would require that one does not collect data right
at threshold, but rather at a point which provides a robust signal
somewhat above threshold.
\par
Equation~(\ref{eq:dMX}) still provides a useful guide in the region
above threshold.  We carried out a Feldman-Cousins confidence-belt
construction, and obtained the statistical uncertainty $\Delta M_X$
as a function of the difference $\sqrt{s}/2-M_X$, {\it i.e.}, the
energy above threshold.  Figure~\ref{f:dMXbelt} shows the result,
based on the simple assumed cross-section $\sigma = \sigma_0\beta^3/s$,
and approximate values for $\eff$ and $\Lum$ corresponding to the
analysis described in Section~\ref{sec:cut}.  As seen in Fig.~\ref{f:dMXbelt}, 
the uncertainty on the mass, $\Delta M_X$, really does decrease as 
$\sqrt{s}/2\rightarrow M_X$,  since the sensitivity of $\sigma(M_X)$ 
to~$M_X$ improves more rapidly than the relative precision on the 
cross-section, $\Delta\sigma/\sigma$, worsens.

\FIGURE{
\includegraphics[width=0.75\textwidth]{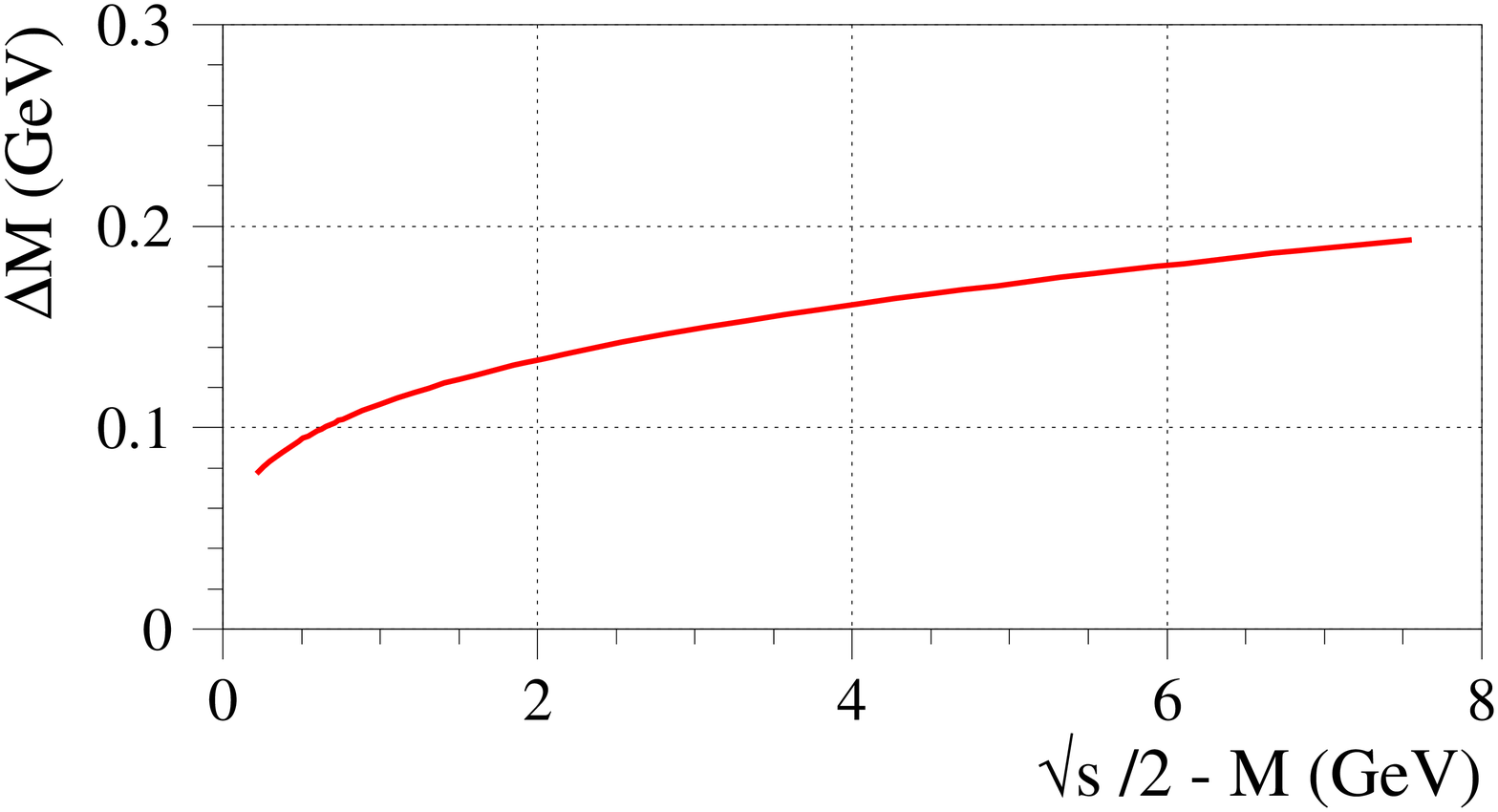}
\caption{\label{f:dMXbelt} Statistical uncertainty on the mass, $\Delta M_X$,
as a function of the beam energy above threshold, $\sqrt{s}/2-M_X$.  This result
is based on a Feldman-Cousins confidence belt construction, with a simple
Ansatz for $\sigma(M_X)$ and approximate values for $\eff$ and $\Lum$.
Backgrounds were not taken into account.}
}


\par
We turn now to a more realistic case.
The theoretical cross section as a function of $\sqrt{s}$ is shown in
Fig.~\ref{cs}, for two stop masses ($122.5\gev$ and $123.5\gev$).
We include QED radiative effects, as described in Section~\ref{sec:th}.
Following the scenario discussed in Ref.~\cite{stop}, we consider
$M_X \approx 123\gev$, and beam polarizations of 80\% for the electron,
and $-60$\% for the positron.
If we want to use a measurement of the cross section to distinguish
these two masses, then clearly the threshold region is the most sensitive.
This corresponds to maximizing $d \sigma/d\, M_X$, which
will minimize $\Delta M_X$ according to Eq.~(\ref{eq:dMX}).
The lower plot on the left side of Fig.~\ref{cs} zooms in on the threshold 
region, to show how much the cross-section differs for two different
hypothetical stop masses, and the lower plot on the right shows
this difference relative to the cross-section for $\mst = 123\gev$.

\FIGURE{
\includegraphics[width=0.95\textwidth]{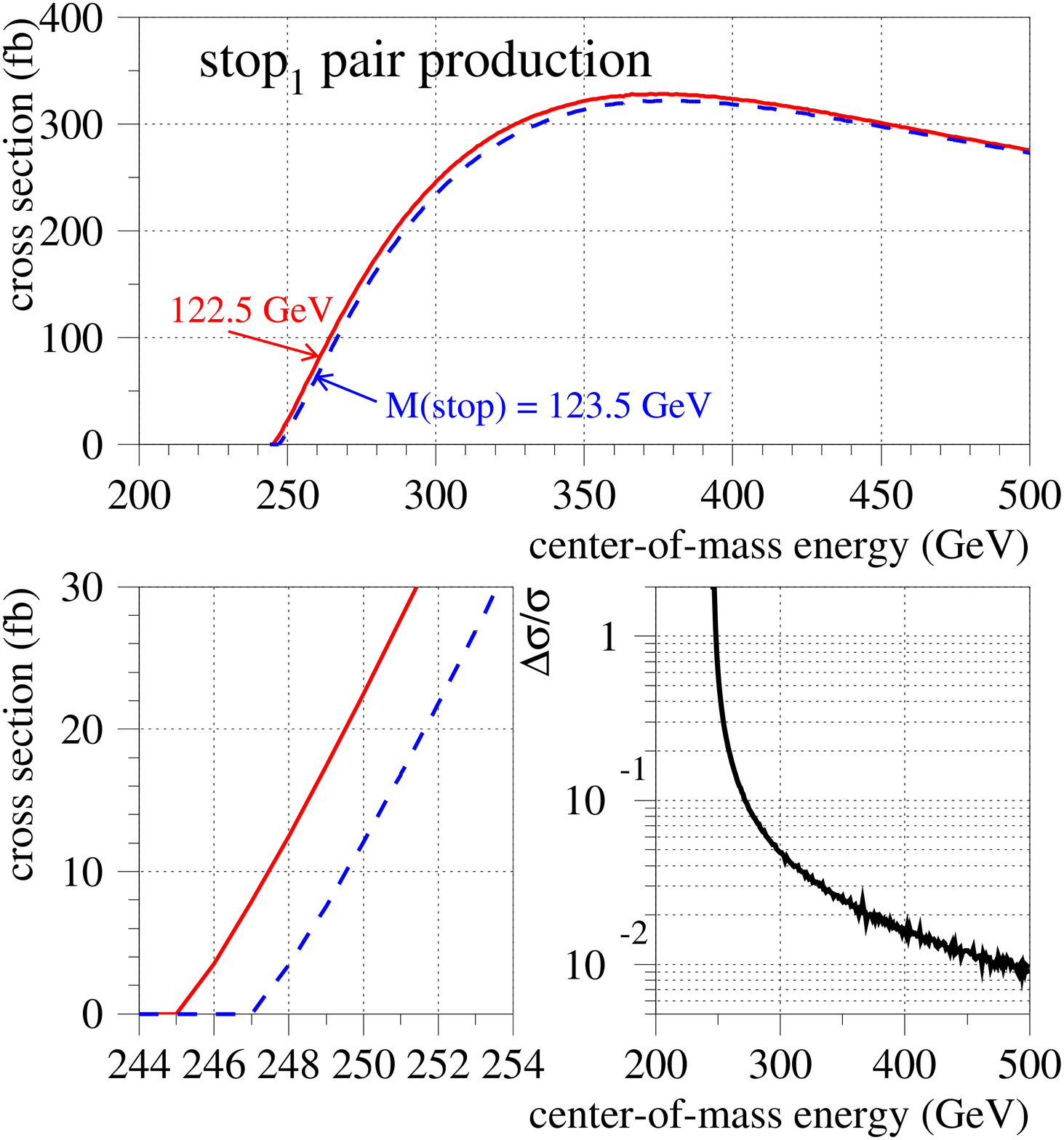}
\caption{\label{cs} Cross-sections for pair production of the 
lightest stop squark.
Top plot shows the full excitation curve as a function of $\sqrt{s}$
for two slightly different values of $\mst$.  The lower left-hand
plot shows a close-up of the threshold region.  The lower right-hand
plot shows the difference of the two cross-sections relative to their
average value.  Clearly the largest relative difference is seen very
close to threshold.}
}

\par
Recall the relation of the cross section to experimental quantities:
\begin{equation}
  \sigma = \frac{\Nev - B}{\epsilon \, \Lum} .
\end{equation}
In a real analysis, $B$, $\epsilon$ and $\Lum$ all carry systematic
uncertainties, which must be assessed and taken into account.
An `optimal' analysis will keep these to a minimum.
\par
Usually the most difficult component in the systematic error
comes from the efficiency and acceptance.  An absolute
cross section requires knowledge of the absolute efficiency, which,
in the case of the $\stopone$ search described in Ref.~\cite{stop},
involves charm-tagging as well as the hadronization and fragmentation
of the $\stopone$ and $c$-quark.  While a large sample of
$\epem\rightarrow\stopone\stoponestar$ events will allow one to
tune Monte Carlo models, and other Standard Model processes may
provide large samples of $c$-jets for measuring efficiencies for
$c$-tagging, it may be useful to have a method which is relatively
insensitive to these sources of systematic uncertainties.
\par
The common step toward reducing systematic uncertainties from
the efficiency is to work with ratios of cross sections.
This also can reduce uncertainties from the luminosity measurement,
and potentially, from the background and theoretical signal cross-section
as well.   We propose to measure the yield of signal
events close to threshold, which will be very sensitive to~$M_X$,
and compare it to the yield near the peak of the excitation curve,
which will be insensitive to~$M_X$ (see Fig.~\ref{cs}).
We define the observable
\begin{equation}
\label{eq:Y}
  Y(M_X,\rtsth) \equiv
  \frac{\Nth-\Bth}{\Npk-\Bpk} = 
  \frac{\sigma_{\st}(\rtsth)}{\sigma_{\st}(\rtspk)} 
  \cdot
  \frac{\effth}{\effpk}
  \cdot
  \frac{\Lumth}{\Lumpk}
\end{equation}
where $\Nth$ and $\Bth$ are the numbers of selected events and
estimated background events for $\rtsth$ near threshold, and $\Npk$, $\Bpk$
are the same quantities for $\rtspk$ near the peak of the
excitation curve.   Anticipating the results of later sections, we 
have computed the observable~$Y$ as a function of~$\mst$, and displayed 
the result in Fig.~\ref{F:y_vs_m}. 
\par
The slope of the line in Fig.~\ref{F:y_vs_m} depends on several
factors, and one can attempt to optimize~$Y$ in order to obtain
the best measurement of~$\mst$.  The sensitivity of $Y$ to $\mst$
comes through $\sigma_{\st}(\rtsth)$, so $\rtsth$ should be close
to $2\mst$, as discussed above.  Mindful of large theoretical and
growing experimental uncertainties as $\rtsth\rightarrow2\mst$,
we have selected $\rtsth = 260\gev$, which is $14\gev$ above the
nominal threshold for a stop with $\mst = 123\gev$.  We find the
peak cross-section occurs at $\rtspk \approx 370\gev$, 
but $\sqrt{s} = 500\gev$ would also serve well.  
Reducing the statistical uncertainty on~$Y$ to an
absolute minimum would require maximizing the integrated luminosity
at threshold, $\Lumth$, but in reality one would not run the ILC
at $\rtsth = 260\gev$ for very long, and in practice $\Lumth = 50~\fbinv$
is already adequate.  We assume $\Lumpk = 200~\fbinv$.
\par
We computed the cross sections with the program {\sc Calvin}~\cite{calvin},
which includes next-to-leading (NLO) order supersymmetric QCD corrections, and 
which was modified for this work to include resummed Coulomb corrections 
near threshold (see Section~\ref{sec:th}).  For two common choices of 
beam polarization, the cross-sections are
\begin{equation}
\begin{aligned}
&P(e^-) = -80\% / P(e^+) = +60\%: & \sigma(\rtsth) &= 17.4~\fb &
	\sigma(\rtspk) &= 72~\fb, \\
&P(e^-) = +80\% / P(e^+) = -60\%: & \sigma(\rtsth) &= 77~\fb &
	\sigma(\rtspk) &= 276~\fb, \\
\end{aligned}
\end{equation}
where $P<0$ stands for left-handed polarization and $P>0$ for right-handed
polarization. We choose the second set of polarization values since it
leads to a much better signal-to-background ratio.
\par
For the computation of the observable $Y$ depicted in Fig.~\ref{F:y_vs_m},
we employed the results of the ``cut-based'' analysis described in
Section~\ref{sec:cut}.  The efficiencies at threshold and peak are 
$\effth = 0.34$ and $\effpk = 0.21$, and the total background cross-sections 
are $2.5~\fb$ and~$10.3~\fb$, respectively.    The strong variation 
of~$Y$ with~$\mst$ in Fig.~\ref{F:y_vs_m} indicates that a precise 
measurement of~$Y$ will lead to a precise value for~$\mst$.  The shaded 
horizontal band corresponds to a~3\% uncertainty on~$Y$, resulting in
$\Dmst = 0.2\gev$, which would be far better
than the result reported in Ref.~\cite{stop}.

\FIGURE{
\includegraphics[width=0.65\textwidth]{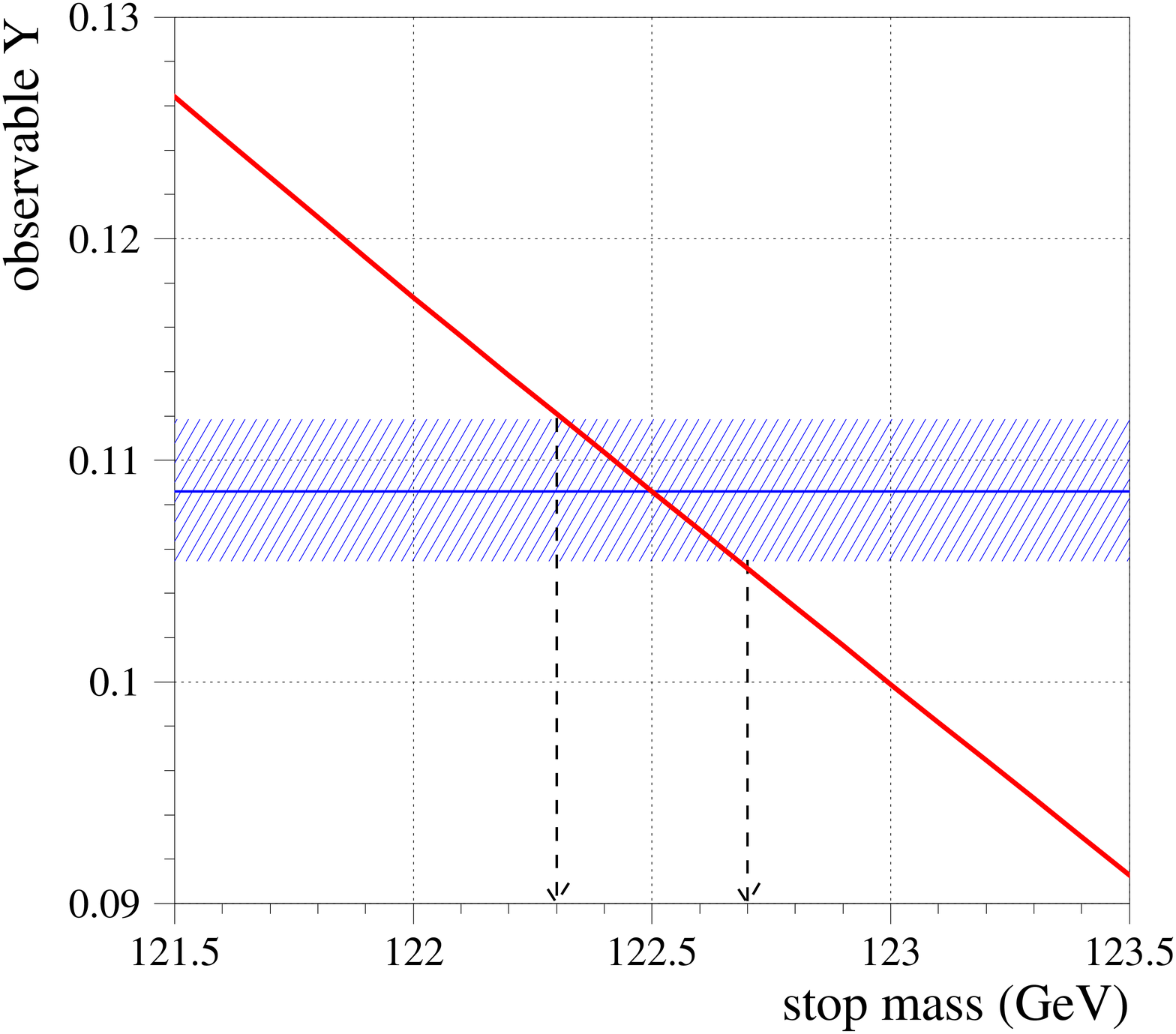}
\caption{\label{F:y_vs_m}
Variation of observable $Y$ with~$\mst$, shown
by the solid red line.  The horizontal line gives the expected
value of~$Y$ when $\mst = 122.5\gev$, and the shaded band shows
a variation of~$Y$ by~3\%.  The vertical arrows indicate
the corresponding uncertainty on~$\mst$.}
}

\par
We consider now the impact of systematic uncertainties on the observable~$Y$,
and Eq.~(\ref{eq:Y}) provides our starting point.  For the event selection
criteria described in Sections~\ref{sec:cut} and~\ref{sec:ida}, the signal
is much bigger than the background, so the main experimental uncertainties 
will come from~$\eff$.  The values for~$\eff$ at threshold and on the peak
come from Monte Carlo simulations of the signal process.  Systematic errors
arise when these simulations do not match reality perfectly.  For example,
the calibration of the calorimeter energy measurement for real data may
be slightly different than is simulated, in which case the efficiency for
a cut on the total visible energy~$\Evis$ as estimated from the simulation 
will be slightly incorrect.  One can express the impact of this error on
the efficiency as $\efftrue = \effest(1+\delta)$, so that $\delta$ is the
{\it relative} shift in the efficiency.  Then the impact on the observable~$Y$
is simply
$$
 \Ytrue = \Yest \left( \frac{1+\deltath}{1+\deltapk} \right) 
 \qquad
 \frac{\Ytrue-\Yest}{\Ytrue}  \approx  \deltapk - \deltath .
$$
Thus, if the systematic uncertainties $\deltapk$ and $\deltath$ are
correlated, and if they have the same relative impact on~$\effpk$
and~$\effth$, the net effect on~$Y$ will be zero, and there
will be no error on~$\mst$.   For some systematic effects, the
errors will be correlated, but of a different magnitude at the
two energies, so that the cancellation $|\deltapk-\deltath|$ will
not be complete.  For other systematic effects, the errors
will be uncorrelated, in which case there is no cancellation.
Clearly the analysis should be designed in such a way as to take
advantage of this cancellation.  In practice, this means that the
cuts should have a similar impact on the signal at both energies.
For the present application, there is a large degree of cancellation,
leading to a greatly reduced systematic uncertainty on the
observable~$Y$, and hence on~$\mst$.  The details are
given in Section~\ref{sec:exp}.
\par
We proceed now to a detailed and realistic simulation, and the
description of two fully-developed event selection methods.


\section{Event Selection and Analysis}
\label{sec:ana}
\par
At an $\epem$ collider, scalar top quarks would be produced in pairs,
and decay to a $c$-quark and the lightest neutralino:
\begin{equation}
e^+e^- \to \tilde{t}_1 \, \tilde{t}_1^* \to c \neu_1 \, \bar{c} \neu_1.
\label{eq:signal}
\end{equation}
The stop quarks live long enough to hadronize before decaying, so the final state
signature consists of two charm quark jets, missing energy and possibly additional
jets due to the hadronization process and gluon radiation.

\par
In the following sections, the method described in Section~\ref{sec:method}
will be applied to the theoretical parameter point of Ref.~\cite{heavyq}.
The weak-scale MSSM parameters are 
\begin{equation}
\begin{aligned}
m^2_{\rm\tilde{U}_3} &= -99^2\gev, &\qquad
m_{\rm\tilde{Q}_3} &= 4330\gev, &
m_{\rm\tilde{Q},\tilde{U},\tilde{D},\tilde{L},\tilde{R}_{1,2}} &= 10~{\mathrm{TeV}},
\\
M_1 &= 118.8\gev,  & M_2 &= 225\gev, & |\mu| &= 225\gev,
\\
A_t &= -1100\gev, & m_{\rm A^0} &= 800\gev, &
\phi_\mu &= 0.2, & \tan\beta &= 5.
\end{aligned}
\end{equation}
The corresponding tree-level masses are:
\begin{equation}
\begin{aligned}
  && \mst &= 122.5\gev, 
  &
   m_{\tilde{t}_2} &= 4333\gev,      
\\
  \mneu{1} &= 107.2\gev,
  &
  \mneu{2} &= 170.8\gev,
  & 
   \mneu{3} &= 231.2\gev,
  &
   \mneu{4} &= 297.7\gev,
\\
   && \mcha{1} &= 162.7\gev,         
   &
   \mcha{2} &= 296.2\gev, 
\label{eq:masses}
\end{aligned}
\end{equation}
and the light stop state is almost completely right-chiral,
$\cos\theta_{\tilde{t}} = 0.010$. As a result of the small stop-neutralino
mass difference, the stop almost completely decays through the loop-induced
process into a charm and neutralino, $\tilde{t} \to c \, \neu_1$. Due to the
loop suppression of the decay, the stop is expected to hadronize before decaying.
We have carried out realistic experimental simulations, and will present the 
analysis of relevant systematic effects.

\subsection{Simulation}
\par
Both the signal and background events are generated with {\sc Pythia
6.129}~\cite{pythia}.  The cross-sections for the signal process were
computed with {\sc Calvin}~\cite{calvin} with some improvements
as in Ref.~\cite{slep}.  The relevant background processes have been 
computed by adapting the Monte Carlo
code used in Ref.~\cite{slep} and by {\sc Grace 2.0} \cite{grace}, with
cross-checks with {\sc CompHep 4.4}~\cite{comphep}.  The simulation and
cross-section calculations incorporated beamstrahlung for cold ILC technology as
parameterized in the program  {\sc Circe 1.0} \cite{circe}. Table~\ref{tab:xsec}
summarizes the predicted  signal and background cross-sections. To avoid the
infrared divergence of the two-photon background process, a cut on the minimal
transverse momentum is applied, $\pt > 5$ GeV. Backgrounds from supersymmetric
processes will be discussed below.  Table~\ref{tab:generated} lists the numbers
of events generated and equivalent luminosity based on the cross-sections in
Table~\ref{tab:xsec}.
\par
Hadronization of the final state charm quark and the intermediate stop quark 
are a key issue in this study.  The Lund string fragmentation model
was used together with the Peterson fragmentation function~\cite{fragfunc}. 
The stop fragmentation is simulated \cite{pythiastops} by labeling the stop quark
as a stable particle in an intermediate step, and switching on the stop decay
again after stop fragmentation. The modeling of the hadronization spectrum of
the stop is described in Ref.~\cite{kraan}. The dominant lightest stop hadron
states are mesons composed of a stop and an up or down quark.
\par
The {\sc Simdet} detector simulation \cite{simdet} was
used, describing a typical ILC detector. The analysis used the 
{\sc N-Tuple} tool~\cite{ntuple}, which incorporates jet-finding algorithms.  
In order to reduce the size of the ntuples, several pre-selection cuts
were applied, as was done for the previous analysis~\cite{stop}:
\begin{eqnarray}
 4 < \Ntracks < 50,   \qquad   \pt > 5\gev,  \cr
 |\costhetathrust| < 0.8,   \qquad
 |p_L / p_{\mathrm{tot}}| < 0.9,  \cr
 \Evis < 0.75\sqrt{s}, \qquad   m_{\mathrm{inv}} < 200 \gev .
\label{ntuplecuts}
\end{eqnarray}
Most of these cuts have very little impact on the signal efficiency.

\TABLE{
\begin{tabular}{|l|rrr|rrr|}
\hline
Process &  \multicolumn{3}{c|}{Cross-section [pb] at $\rtsth = 260 \gev$} 
       &  \multicolumn{3}{c|}{Cross-section [pb] at $\rtspk = 500 \gev$} \\
\hline
$P(e^-) / P(e^+)$ & 0/0 & $-$80\%/+60\% & +80\%/$-$60\%
		  & 0/0 & $-$80\%/+60\% & +80\%/$-$60\% \\
\hline
$\tilde{t}_1 \tilde{t}_1^*$ & 0.032 & 0.017 & 0.077 & 0.118 & 0.072 & 0.276 \\
\hline
$W^+W^-$ & 16.9\phantom{0} & 48.6\phantom{0} & 1.77 
         & 8.6\phantom{0} & 24.5\phantom{0} & 0.77 \\
$ZZ$    & 1.12 & 2.28 & 0.99 & 0.49 & 1.02 & 0.44 \\
$W e\nu$ & 1.73 & 3.04 & 0.50 & 6.14 & 10.6\phantom{0} & 1.82 \\
$e e Z$  & 5.1\phantom{0} & 6.0\phantom{0} 
	 & 4.3\phantom{0} & 7.5\phantom{0} & 8.5\phantom{0} & 6.2\phantom{0} \\
$q \bar{q}$, $q \neq t$ & 49.5\phantom{0} & 92.7\phantom{0} & 53.1\phantom{0} 
			& 13.1\phantom{0} & 25.4\phantom{0} & 14.9\phantom{0} \\
$t \bar{t}$ & 0.0\phantom{0} & 0.0\phantom{0} & 0.0\phantom{0} & 0.55 & 1.13 &
0.50 \\
2-photon & 786\phantom{.00}&&&  936\phantom{.00}&& \\[-1ex]
$\quad \pt > 5$ GeV &&&&&& \\
\hline
\end{tabular}
\caption{Cross-sections for the stop signal and Standard Model background
processes for $\rtsth = 260 \gev$ and $\rtspk = 500 \gev$ and different
combinations of beam polarization. The signal is given for a right-chiral stop 
of $\mst = 122.5\gev$. Negative polarization values refer to
left-handed polarization and positive values to right-handed polarization.
\label{tab:xsec}}
}

\TABLE{
\begin{tabular}{|l|r|rr|r|rr|}
\hline
 &  \multicolumn{3}{c|}{$\rtsth = 260\gev$} 
 &  \multicolumn{3}{c|}{$\rtspk = 500\gev$}\\
\hline
                  & generated & \multicolumn{2}{c|}{luminosity ($\fbinv$)}
		  & generated & \multicolumn{2}{c|}{luminosity ($\fbinv$)} \\
\hline
$P(e^-) / P(e^+)$ &   & 0/0 & {+80\%/$-$60\%}
		  &   & 0/0 & {+80\%/$-$60\%} \\
\hline
$\tilde{t}_1 \tilde{t}_1^*$ & 50,000 & 1562 & 649
                            & 50,000 &  423 & 181  \\
\hline
$W^+W^-$ & 180,000 &     11 &   102 &   210,000 &      24 &    273 \\
$ZZ$     &  30,000 &     27 &    30 &    30,000 &      61 &     68 \\
$W e\nu$ & 210,000 &    121 &   420 &   210,000 &      34 &    115 \\
$e e Z$  & 210,000 &     41 &    49 &   210,000 &      28 &     34 \\ 
$q \bar{q}$, $q \neq t$ 
         & 350,000 &      7 &     6 &   350,000 &      27 &     23 \\
$t \bar{t}$ & ---  &   ---  &   --- &   180,000 &     327 &    360 \\
2-photon & $1.6\times 10^6$ 
         &   2    &  2   & $8.5\times 10^6$ &  9 &  9 \\
\hline
\end{tabular}
\caption{ \label{tab:generated}  
Numbers of generated events at $\rtsth = 260\gev$ and $\rtspk = 500\gev$,
and the equivalent luminosities in~$\fbinv$.}
}

\subsection{Sequential-Cut Analysis}
\label{sec:cut}
\par
Although Standard Model background processes are several orders of magnitude
larger than the stop signal process, the background contributions can be reduced to 
an acceptable level by suitable selection cuts. This work follows the analysis of
Ref.~\cite{stop}, but makes some adjustments to accommodate the stop
fragmentation effects, and to take advantage of the cancellation of
systematic uncertainties as discussed in Section~\ref{sec:method}.

The event selection begins with some basic and common kinematic cuts
based on global event quantities.  The visible energy, $\Evis$, must be
less than $0.3\sqrt{s}$ to ensure a large missing-energy signature.  It must
be greater than $0.1\sqrt{s}$ to suppress the bulk of the two-photon
events.  Similarly, the number of reconstructed charged tracks must
indicate real hadronic jets, so we require $\Ntracks  \ge 5$.  In order
to suppress $W e\nu$ and $\qqbar$ signals, we place an upper bound
$\Ntracks \le 25$ at threshold and $\Ntracks \le 20$ at peak.  These
cuts on $\Ntracks$ remove only a couple percent of the signal.

We place one more kinematic and one topological cut to further reduce 
the backgrounds.  The cuts values are carefully tuned to achieve a low
systematic uncertainty for the observable~$Y$, as well as a good
background rejection.  In practice, this means aiming to remove 
approximately the same amount of signal at the two center-of-mass 
energies, rather than achieving the highest signal efficiencies.
In particular, the efficiency at $\rtspk$ is relatively unimportant
since we anticipate a large luminosity and a large signal cross-section
there.  The thrust value is useful for eliminating $\qqbar$ and
two-photon events.  As shown in Fig.~\ref{f:marg_thrust}, the thrust distribution 
for the signal is rather different at the two center-of-mass energies,
so we require $0.77\le T\le 0.97$ at $\rtsth$ and $0.55\le T\le 0.90$
at $\rtspk$.  Similarly, the event~$\pt$, calculated from all energy
flow objects in the event, is crucial for eliminating the two-photon
background.  Our study indicates that a minimum cut $\pt > 15\gev$
is needed.  We tighten this cut to $\pt > 22\gev$ at $\rtspk$
in order to eliminate the same amount of signal events as are 
eliminated at $\rtsth$.
Fig.~\ref{f:marg_pt} shows that cutting at $\pt = 22\gev$ at $\rtspk$ places the
cut at almost the same point in the $\pt$ distribution for both center-of-mass
energies.  An upper cut on~$\pt$ helps reduce the $W e\nu$ background,
so we require $\pt < 45\gev$ at $\rtsth$ and $\pt < 50\gev$ at $\rtspk$,
which again reflects our effort to minimize the systematic uncertainty.

\FIGURE{
\mbox{
\includegraphics[width=0.47\textwidth]{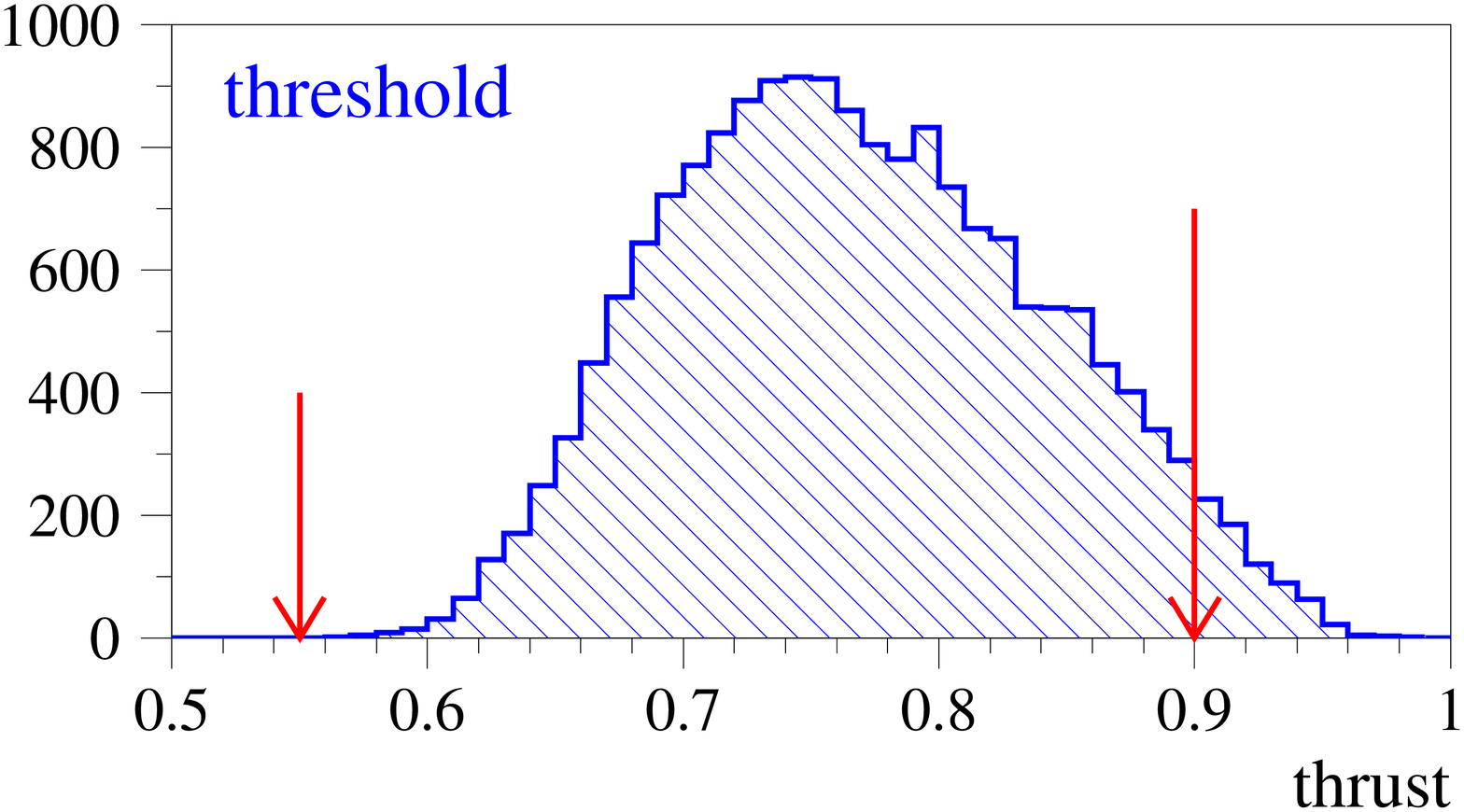}
\includegraphics[width=0.47\textwidth]{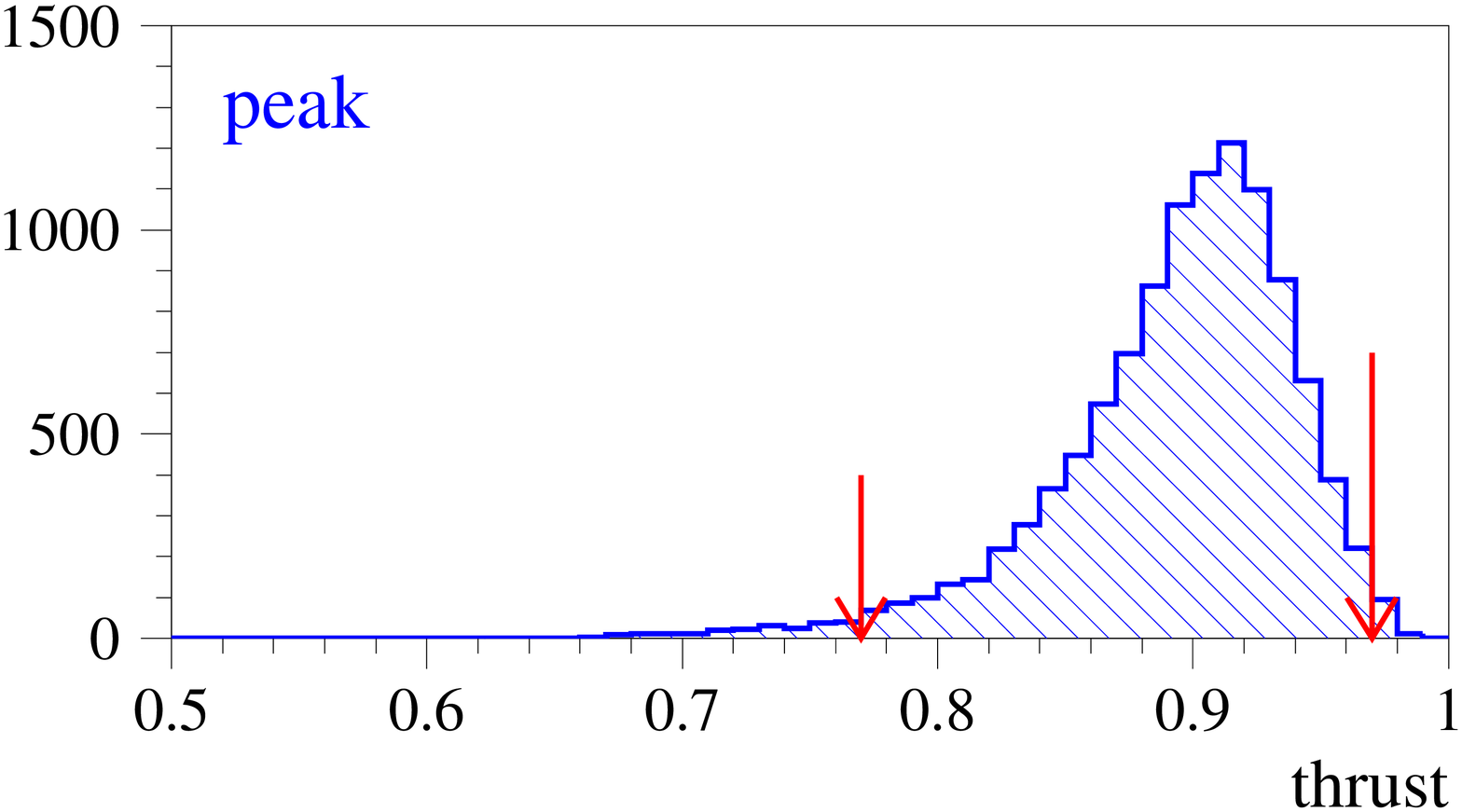}
}
\caption{\label{f:marg_thrust} Signal marginal distributions 
for the thrust, at~$\rtsth$~(left) and at~$\rtspk$~(right).
The arrows indicate the cuts chosen to minimize the systematic
uncertainties as well as the statistical uncertainty.}
}

\FIGURE{
\mbox{
\includegraphics[width=0.49\textwidth]{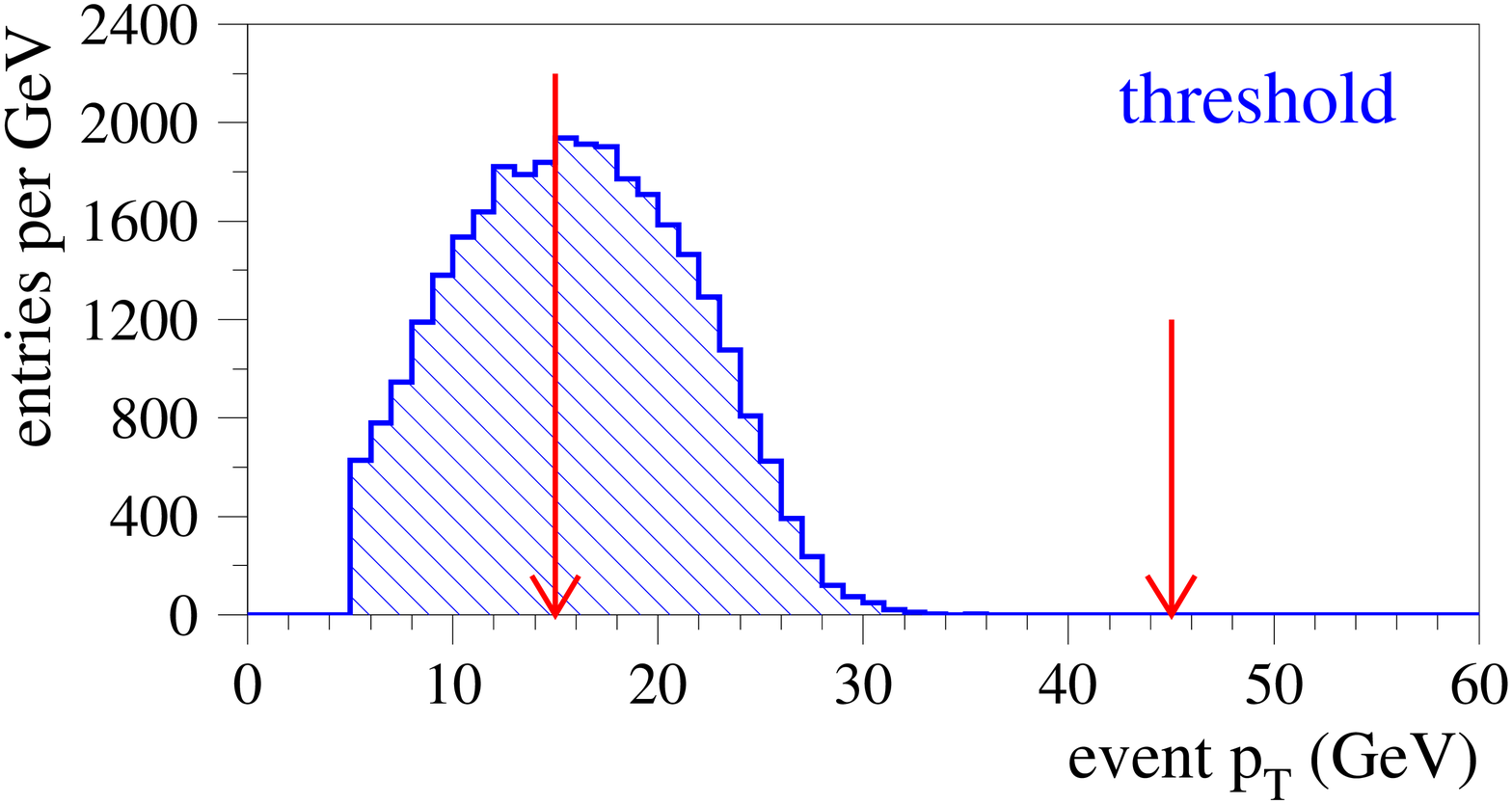}
\includegraphics[width=0.49\textwidth]{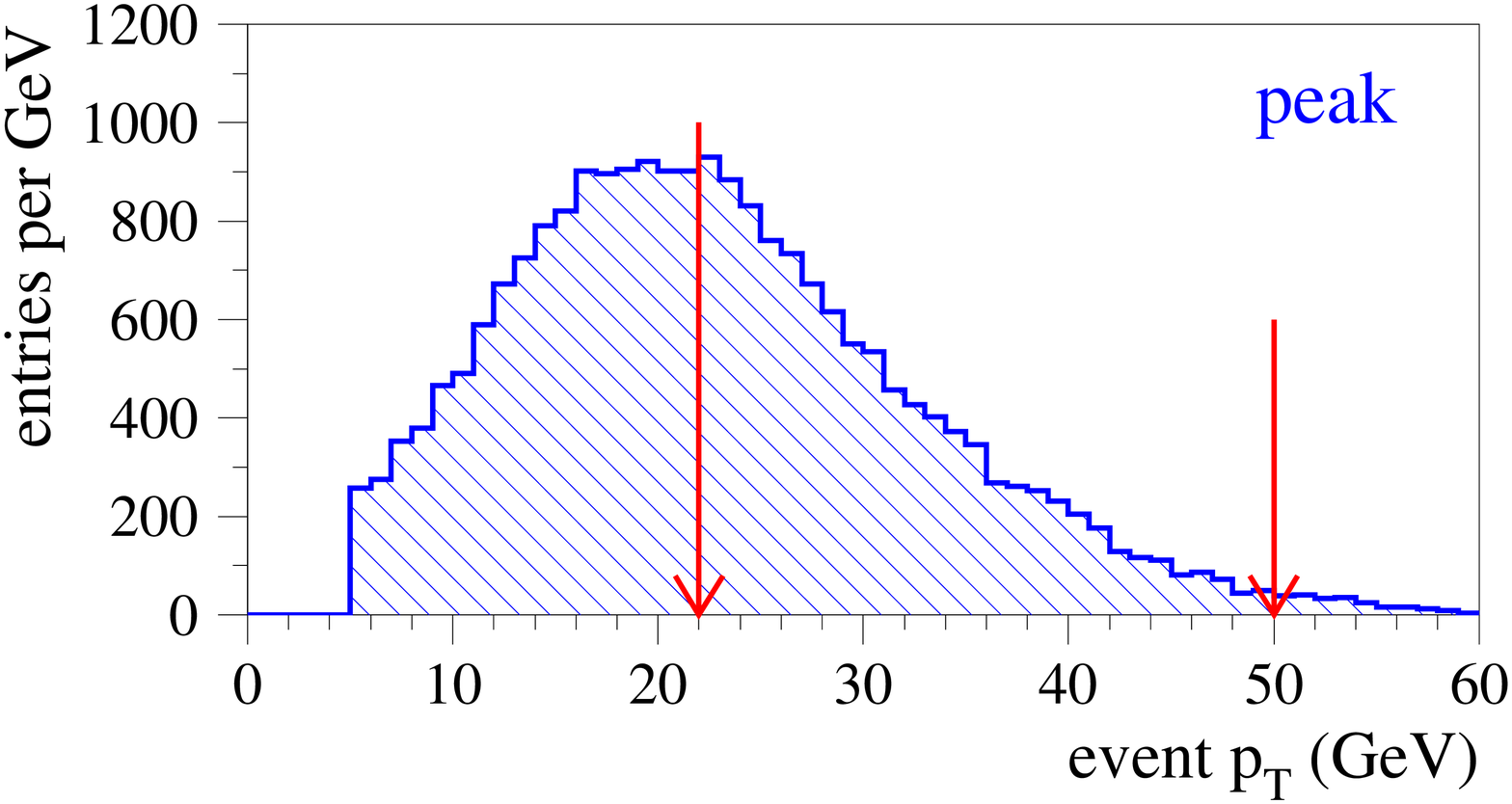}
}
\caption{\label{f:marg_pt} Signal marginal distributions 
for~$\pt$, at~$\rtsth$~(left) and at~$\rtspk$~(right).
The arrows indicate the cuts chosen to minimize the systematic
uncertainties as well as the statistical uncertainty.}
}

\par
One might expect that the signal process~(\ref{eq:signal}) produces only
two jets.  However, additional soft jets can emerge from the stop hadronization
process and also from the decay of the stop hadron.  In order to maintain
a high efficiency, and to avoid large systematic uncertainties from the
modeling of the rate and characteristics of these extra jets, events with 
more than two jets should not be rejected. However, to suppress the background
processes effectively, extra jets are allowed only when their energy falls
below a certain cut-off value.  To be specific, if there are more than two
jets in an event, only two of the jets are allowed to have energies
above~$25\gev$.  In this paper, we refer to this requirement as the
``extra-jet veto.''   Furthermore, if there are more than three jets,
the most energetic jet cannot be too energetic -- its energy must be
less than~$35\gev$.  These cuts are useful against the troublesome
$W e\nu$ backgrounds, especially at~$\rtspk$.
\par
Further substantial improvements of the signal-to-background ratio can be 
achieved by exploiting kinematic and topological correlations between the 
two $c$-quark jets.  Therefore it is necessary to identify them from the 
plurality of jets, and for this we use charm tagging as realized using a
neural network~\cite{kuhlhiggs}. 
The neural network uses information about the vertex position of a jet based on
a topological vertex finder, the impact parameter probability, the momenta of
the associated tracks and the reconstructed mass. It has been 
optimized to single out charm jets with an energy that is typical for light 
stops, while rejecting light quark jets coming from $We\nu$ background.
Each jet in an event is tested with the charm tagger, and a charm flag~$\cflag$
is set (ideally, $\cflag = 1$ indicates a charm jet).  First, if a displaced vertex is 
reconstructed, the jet is tagged positively with $\cflag = 1$.  A displaced
vertex is found roughly 50\% of the time for a charm jet, and less than 20\% of
the time for a light quark jet.  If no such vertex 
is reconstructed, then the neural network is employed, which produces a charm flag 
value between zero and one, $0 \le \cflag \le 1$.  The output of the
neutral network is shown in Fig.~\ref{f:ctag_ann}, for the second of
the two charm-tagged jets.
\par
We consider the two jets in the event with the highest values of~$\cflag$, 
and require $\Pcharm \equiv \cflagone\times\cflagtwo > 0.6$, 
which is very effective at eliminating events with no charm-quark jets while
retaining a high efficiency for signal events.  In particular, the
$W e\nu$ background is reduced by more than half.  Fig.~\ref{f:ctag_product}
compares the quantity $\Pcharm$ for signal events and $We\nu$ background
which have passed the kinematic event selection cuts.  Since half of
the $We\nu$ events have a genuine charm jet, it is the value of $\cflag$
for the second jet which best distinguishes signal and background.  
\par
A further substantial  reduction can be obtained from cuts on the invariant 
mass of the the two best charm-tagged jets -- we veto events in which that mass is
consistent with the $W$-boson mass.

\FIGURE{
\includegraphics[width=0.65\textwidth]{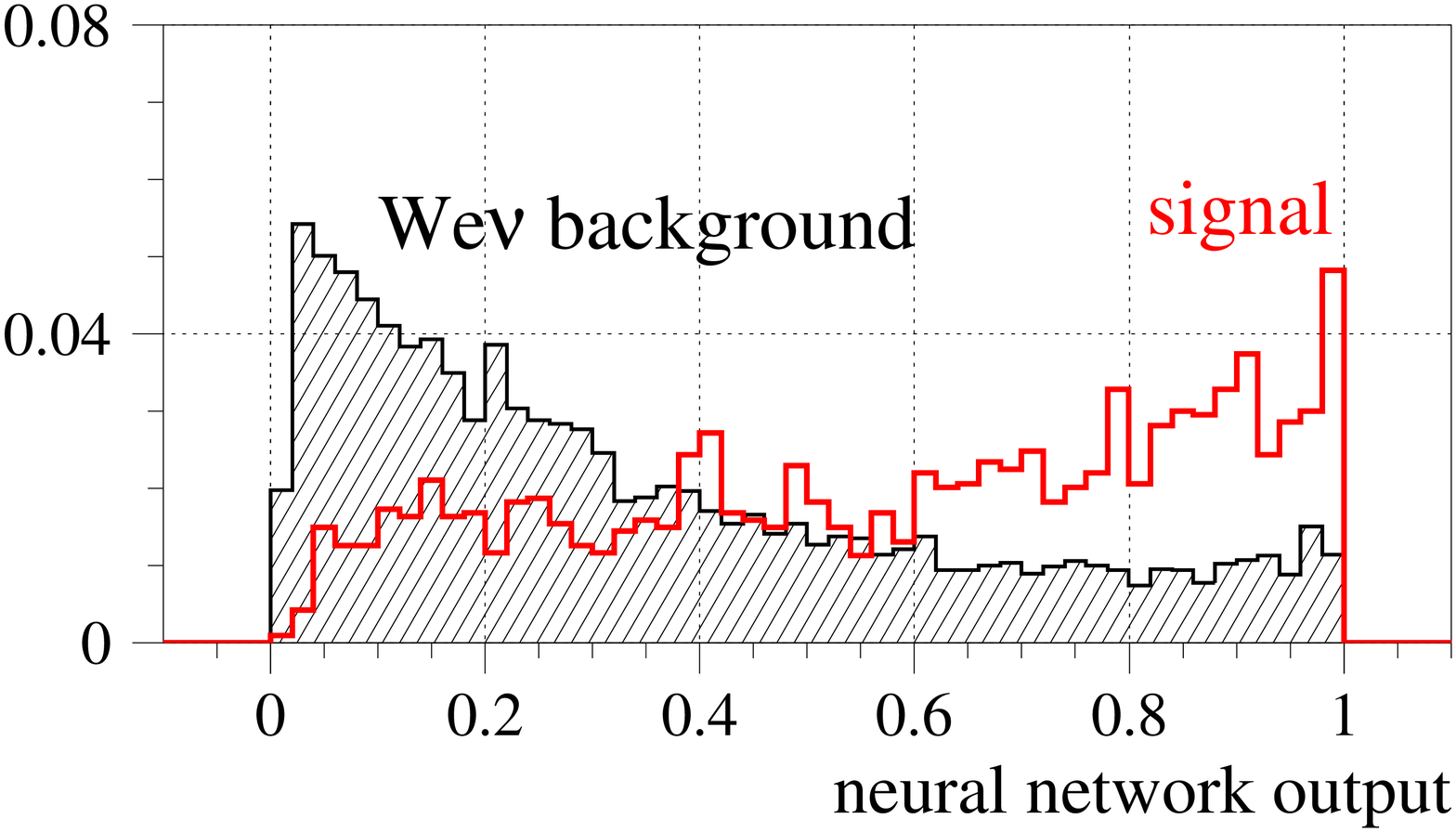}
\caption{\label{f:ctag_ann} Illustration of the ability of
the neural network to discriminate signal and
the main background coming from $We\nu$ production,
for the second of two charm-tagged jets.
Both distributions are normalized to unit area.}
}

\FIGURE{
\includegraphics[width=0.65\textwidth]{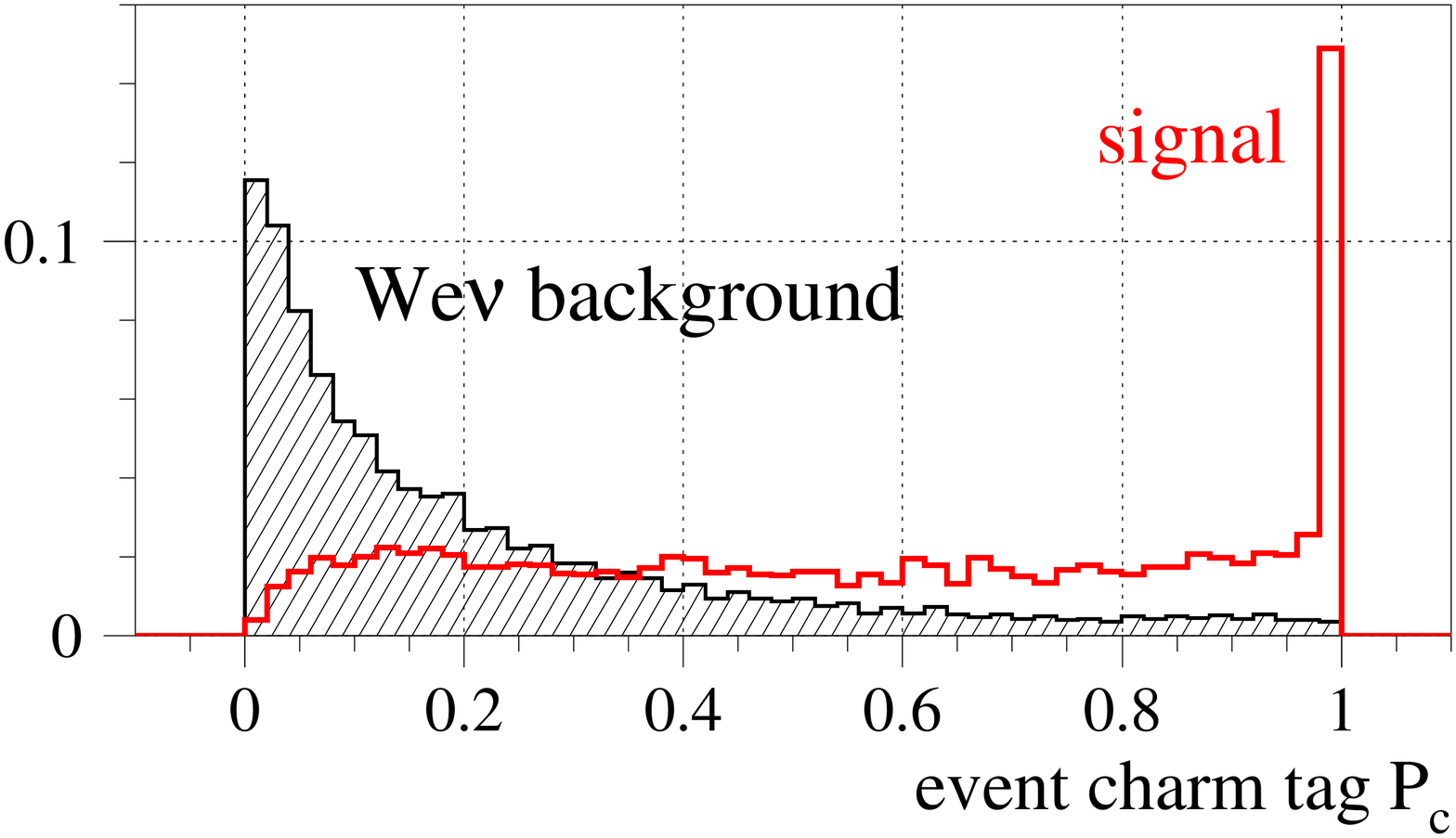}
\caption{\label{f:ctag_product} Event charm probability,
$\Pcharm = \cflagone\times\cflagtwo$, comparing signal
and the $We\nu$ background.  Our requirement is $\Pcharm > 0.6$.
These distributions are normalized to unit area.}
}

\par
The event selection cuts are summarized in Table~\ref{tab:cuts}, 
for the two center-of-mass energies, $\rtsth = 260\gev$ and 
$\rtspk = 500\gev$.   These follow the pre-selection cuts listed 
in Eq.~(\ref{ntuplecuts}).

\TABLE{
\begin{tabular}{|l|c|c|}
\hline
Variable &  $\rtsth = 260 \gev$ 
       &  $\rtspk = 500 \gev$ \\
\hline
number of charged tracks      
   &  $5 \le \Ntracks \le 25$ 
   &  $5 \le \Ntracks \le 20$ 
\\
visible energy $\Evis$
   &  $0.1 < \Evis/\sqrt{s} < 0.3$
   &  $0.1 < \Evis/\sqrt{s} < 0.3$
\\
event longitudinal momentum
   &  $ |p_L / p_{\mathrm{tot}}| < 0.85$
   &  $ |p_L / p_{\mathrm{tot}}| < 0.85$
\\
event transverse momentum $\pt$
   &  $15 < \pt < 45$~GeV
   &  $22 < \pt < 50$~GeV
\\
thrust $T$
   &  $0.77 < T < 0.97$
   &  $0.55 < T < 0.90$
\\
number of jets $\Njets$ 
   &  $\Njets \ge 2$ 
   &  $\Njets \ge 2$
\\
extra-jet veto
   &  $E_{\mathrm{jet}} < 25$~GeV
   &  $E_{\mathrm{jet}} < 25$~GeV
\\
charm tagging likelihood $\Pcharm$ 
   & $\Pcharm> 0.6$ 
   & $\Pcharm> 0.6$
\\
di-jet invariant mass $\Mjj$ 
   & $\Mjjsq < 5500 \gevsq$ or 
   & $\Mjjsq < 5500 \gevsq$ or 
\\

   & $\Mjjsq > 8000 \gevsq$
   & $\Mjjsq > 10000 \gevsq$
\\
\hline\hline
signal efficiency & 0.340 & 0.212 \\
\hline
\end{tabular}
\caption{Selection cuts for $\rtsth = 260 \gev$ and $\rtspk = 500 \gev$.
Also listed are the selection efficiencies for right-chiral stop squarks
and neutralinos with masses given in Eq.~(\ref{eq:masses}).
See the text for explanations of the extra-jet veto, charm tagging,
and the $\Mjjsq$ cut. \label{tab:cuts} }
}

\TABLE{
\begin{tabular}{|l|rr|rr|}
\hline
 &  \multicolumn{2}{c|}{$\rtsth = 260\gev$} 
 &  \multicolumn{2}{c|}{$\rtspk = 500\gev$}\\
\hline
                  & \multicolumn{2}{c|}{$\Lum = 50~\fbinv$}
		  & \multicolumn{2}{c|}{$\Lum = 200~\fbinv$} \\
\hline
$P(e^-) / P(e^+)$    & 0/0 & {+80\%/$-$60\%}
		     & 0/0 & {+80\%/$-$60\%} \\
\hline
$\tilde{t}_1 \tilde{t}_1^*$ & 544 & 1309 
                            & 5170 & 12093  \\
\hline
$W^+W^-$ &   $38$ &   $4$ &    $16$ &    $2$  \\
$ZZ$     &    $8$ &   $7$ &    $36$ &   $32$ \\
$W e\nu$ &  $208$ &  $60$ &  $7416$ & $2198$ \\
$e e Z$  &    $2$ &   $2$ &   $< 7$ &  $< 6$ \\ 
$q \bar{q}$, $q \neq t$ 
         &   $42$ &  $45$ &    $15$ &   $17$ \\
$t \bar{t}$ & $0$ &   $0$ &     $7$ &    $7$ \\
2-photon  &  $53$ &  $53$ &    $12$ &   $12$  \\
\hline
total background & $351$ & $171$ & $7509$ & $2274$ \\
\hline
$S / B$          & $1.5$ & $7.6$ & $0.7$ & $5.3$ \\
\hline
\end{tabular}
\caption{ \label{tab:nev}  
Expected numbers of events
remaining at $\rtsth = 260\gev$ and $\rtspk = 500\gev$, with 
unpolarized and with polarized beams,
after sequential selection cuts have been applied.
The entries in the form $< N$ show the number of events
corresponding to a single selected simulated event.}
}

Our estimates of the numbers of signal and background events
surviving the cuts listed in Table~\ref{tab:cuts} are summarized
in Table~\ref{tab:nev}.   If, in a given channel, no simulated
events remain after applying our cuts, we list an upper limit
corresponding to one simulated event, and we count this amount
in the total background estimate.   As evident from the table, 
the background can be greatly reduced for $\rtsth$, 
resulting in a very good signal-to-background ratio.  At $\rtspk$, 
on the other hand, a large background from single-$W$ boson production is left.
For unpolarized beams, the resulting signal-to-background ratio is~$0.7$
While this would allow an unambiguous discovery of stop 
quarks (see Section~\ref{S:discovery}),   it is not a very good basis for 
precision measurements of the stop mass.
Fortunately, the signal-to-background ratio can be greatly improved by using 
polarized beams.  With an essentially right-handed electron beam and 
left-handed positron beam, the signal is enhanced, while most backgrounds 
are substantially suppressed. As a result, the signal-to-background ratio 
at $\rtspk = 500\gev$ is improved from~$0.7$ to~$5.3$.

We checked for possible supersymmetric backgrounds.  The main concern is
chargino pair production with the decay
channel $\tilde{\chi}_1^+ \rightarrow \sta b$.  We simulated a sample
of these decays, consistent with our benchmark scenario, and found
that the cuts listed in
Table~\ref{tab:cuts} completely eliminate this background source.

With the results listed in Table~\ref{tab:nev} for polarized beams,
we can compute the observable~$Y$ and its statistical error,
obtaining $Y = 0.1082\pm 0.0034$ with a relative error of~3.1\%.
The corresponding stop quark mass would be
\begin{equation}
     \mst = (122.5 \pm 0.19) \gev
\label{staterrcuts}
\end{equation}
where the uncertainty depends on the slope, $dY/dM = -0.01755$,
at $Y = 0.1082$.
Without positron polarization, $P(e^+) = 0$, the precision of the 
measurement is reduced by roughly 20\%, resulting in $\Delta Y / Y = 3.7$\% 
and $\Delta\mst = 0.23\gev$.
Even in this case the statistical error is rather small.

It should be recalled that the production cross-section is a strong
function of the mixing angle, so the statistical error $\Delta\mst$
will also depend on it.   In our reference scenario, the light stop eigenstate 
is almost completely composed of the partner of the right-handed stop, 
$\st_1 \approx \st_{\rm R}$, with the mixing angle $\cos\theta_{\tilde{t}} = 0.01$. 
While this scenario is preferred by electroweak precision data and
the explanation of baryogenesis, an experimental analysis should consider all
possible values for the stop mixing angle.
For other values of $\cos\theta_{\tilde{t}}$, the production
cross-section can change drastically, depending on the beam polarization. As
concrete example, we consider two larger values of $\cos\theta_{\tilde{t}}$:
\begin{align}
\cos\theta_{\tilde{t}} &= 0.6: &
\sigma_{\rm L,260} &= 52 \mbox{ fb}, &
\sigma_{\rm L,500} &= 194 \mbox{ fb}, \\
&& 
\sigma_{\rm R,260} &= 39 \mbox{ fb}, &
\sigma_{\rm R,500} &= 148 \mbox{ fb}, \nonumber \\[1ex]
\cos\theta_{\tilde{t}} &= 1.0: &
\sigma_{\rm L,260} &= 169 \mbox{ fb}, &
\sigma_{\rm L,500} &= 577 \mbox{ fb}, \\
&& 
\sigma_{\rm R,260} &= 6.9 \mbox{ fb}, &
\sigma_{\rm R,500} &= 30 \mbox{ fb}. \nonumber 
\end{align}
Here $\sigma_{{\rm L/R},E}$ stands for the stop production cross-section at
center-of-mass energy~$E\gev$, and with beam polarization combinations 
$P(e^-) = -80\% /P(e^+) = +60\%$ and $P(e^-) = +80\% / P(e^+) = -60\%$, 
respectively. If the stop is dominantly left-chiral, with
$|\cos\theta_{\tilde{t}}| > 0.5$, the production cross-section is substantially
larger for left-handed electron and right-handed positron polarization, opposite
to the situation for a right-chiral stop. Therefore, for large values of
$|\cos\theta_{\tilde{t}}|$, it is better to use the beam polarizations 
$P(e^-) = -80\% / P(e^+) = +60\%$, even though one has to deal with much larger 
Standard Model background. The largest background, $e^+e^- \to W e \nu$, amounts 
to about $12800$~events at $\sqrt{s} = 500\gev$ and $\Lum = 200~\fbinv$ for this 
polarization.  Nevertheless, due to large signal cross-sections, the resulting
statistical error is still small, as summarized in Table~\ref{tab:mixang},
which demonstrates that, for all values of the stop mixing angle, one can
measure the stop mass with a statistical error better than~$0.3\gev$ 
using our method and an appropriate choice of beam polarization.
\TABLE{
\begin{tabular}{|l|c|c|}
\hline
$\cos\theta_{\tilde{t}}$
 & $P(e^-) = -80\% / P(e^+)
= +60\%
$ & $P(e^-) = +80\% / P(e^+)
= -60\%
$ \\
\hline
0.0 & 0.69 & {\bf 0.19} \\
0.6 & 0.29 & {\bf 0.28} \\
1.0 & {\bf 0.14} & 0.94 \\
\hline
\end{tabular}
\caption{Statistical uncertainties $\Delta\mst$ in $\gev$, for selected 
values of $\cos\theta_{\tilde{t}}$ and two opposite sets of beam polarization. 
The bold numbers indicate the best choice of beam polarization for the given 
value of the stop mixing angle.
\label{tab:mixang}}
}

\subsection{Iterative Discriminant Analysis}
\label{sec:ida}
\par
A traditional, sequential-cut analysis was presented in the
previous section.  Often, more advanced multi-variate techniques
can boost the sensitivity of a search.  We investigated the
efficacy of an Iterative Discriminant Analysis~(IDA) for the
purposes of measuring the stop quark mass based on the
observable~$Y$.

The IDA method~\cite{ida} is a modified Fisher Discriminant Analysis, 
the two main differences are the introduction of a non-linear
discriminant function and iterations in order to enhance the separation 
of signal and background. Two IDA steps have been performed.
In order to have two independent samples for the derivation 
of the IDA function and for the expected performance, the signal and 
background samples were divided into two equally-sized samples.  
For this analysis, the same kinematic variables 
and simulated event samples as in the cut-based analysis are used,
including the charm tagging flags $\cflag$. Before the multi-variable 
analysis is performed, cuts on the input variables, so-call ``tail cuts,''
are applied in order to improve the IDA performance. This is achieved by
reducing the number of input events, and thus reducing the computational
time. From the distributions of the input variables for the signal and 
background events, the IDA method calculates a separating surface 
in the multi-dimensional parameter space between signal and
background events. The IDA output variable has a different shape for signal and
background events, and therefore a cut on this variables is used to
separate signal and background. In the first IDA step a cut is placed
on this IDA output variable such that 99.5\% of the signal efficiency are
kept. The number of background events is largely reduced. From the smaller
background sample and the 99.5\% remaining signal events again a new IDA output
variable is constructed. The cut on the IDA output variable in this second IDA
step defines the signal efficiency and the corresponding number of background
events. Different working points are possible: they are defined by choosing a
certain signal efficiency and obtaining the corresponding number of background
events. The working point was determined by the expected error on~$\mst$.
The results of the IDA method with stop fragmentation are shown
in Fig.~\ref{fig:ida2} and Table~\ref{tab:ida2} expressed as number of 
expected background events for each contributing background process.

\FIGURE{
{\bf (a)} $\sqrt{s} = 260 \gev$ \hspace{4.8cm}
{\bf (b)} $\sqrt{s} = 500 \gev$\\[-2em]
\mbox{
\epsfig{figure=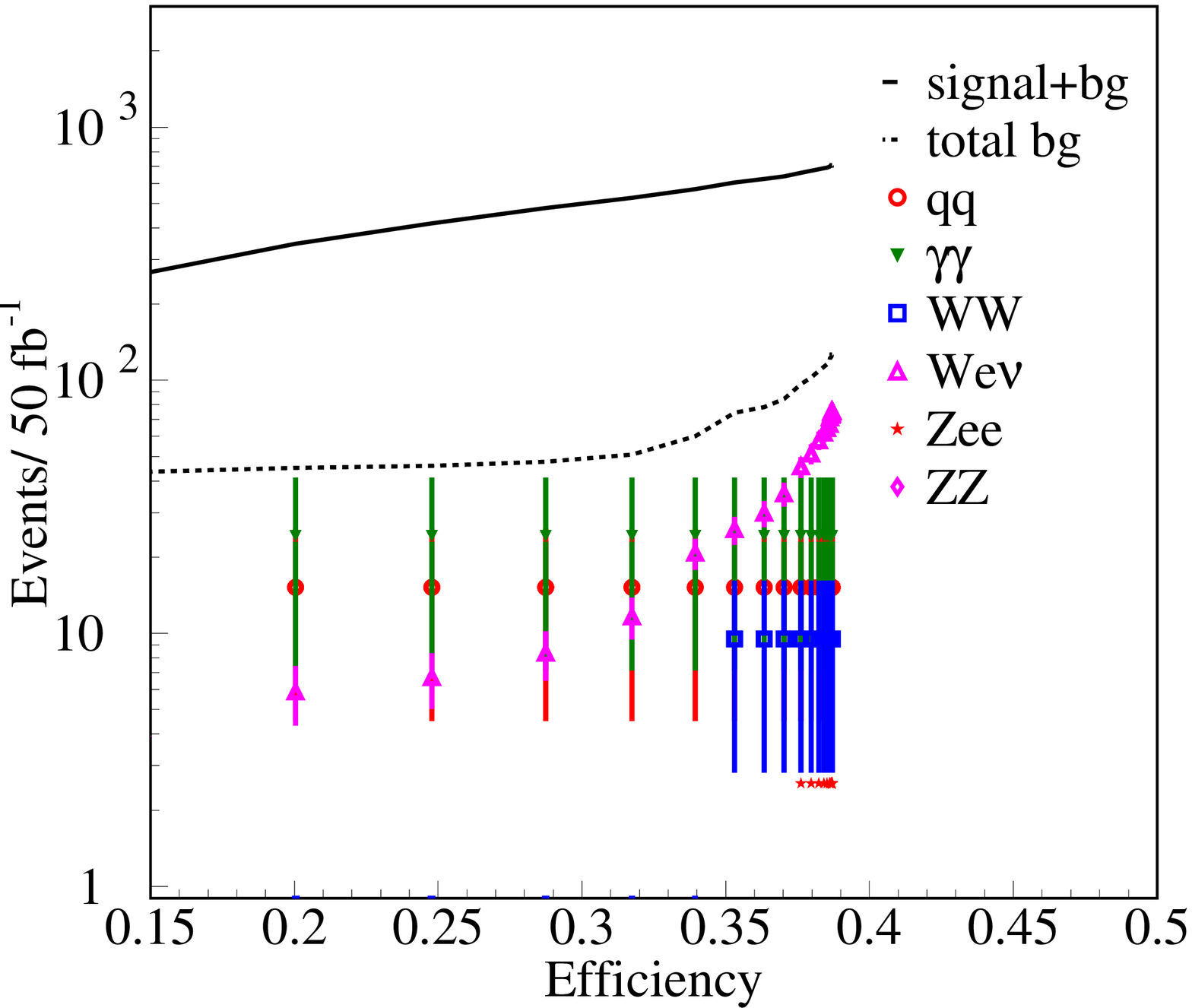, width=8cm}
\epsfig{figure=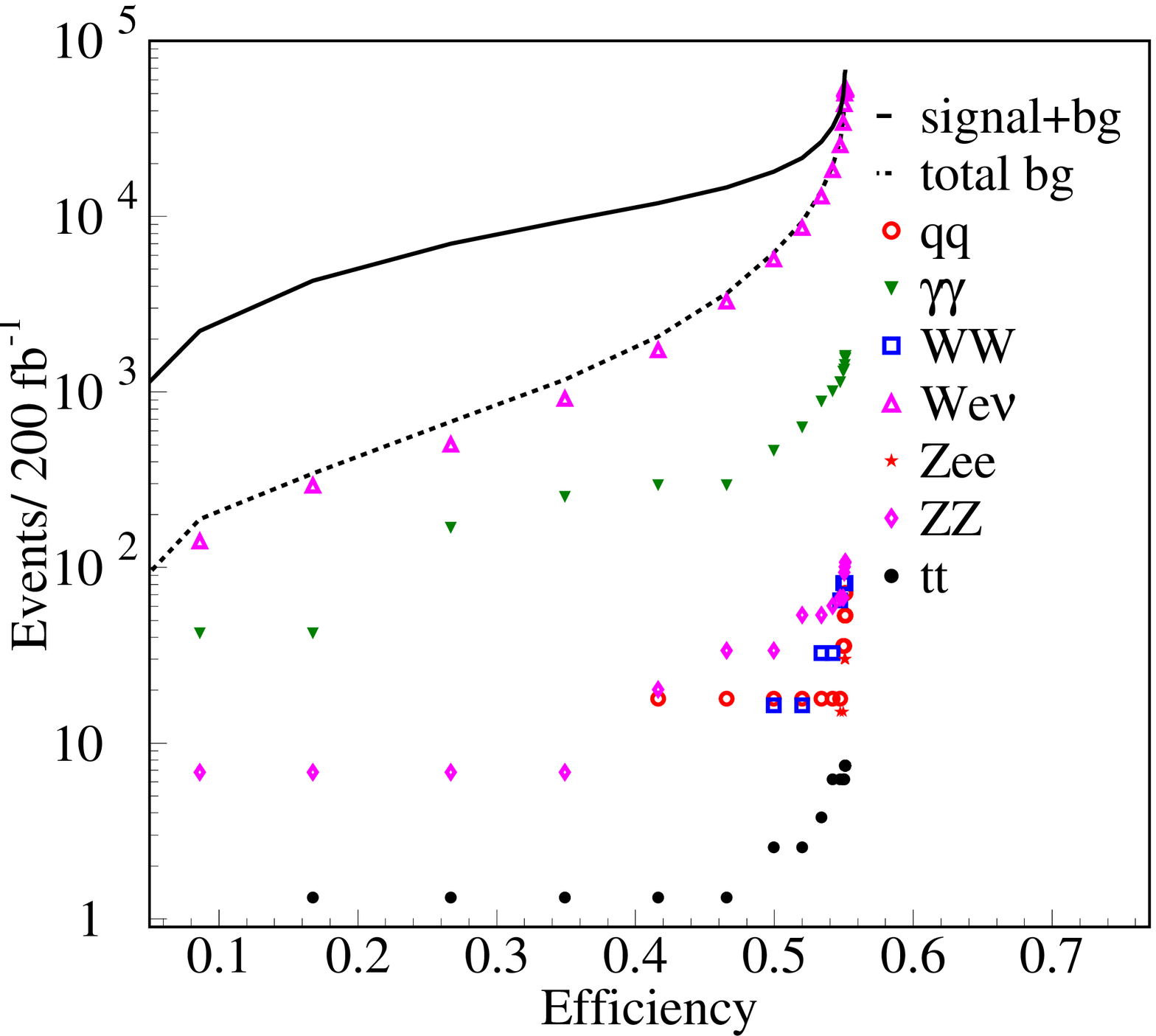, width=8cm}
}
\caption{Performance of the Iterative Discriminant Analysis (IDA) for
$\sqrt{s} = 260 \gev$ and $\sqrt{s} = 500 \gev$.
The plots show the remaining
background event numbers for unpolarized beams as a function of the signal
efficiency.
\label{fig:ida2}}
}

\TABLE{
\begin{tabular}{|l|rr|rr|}
\hline
 &  \multicolumn{2}{c|}{$\rtsth = 260\gev$} 
 &  \multicolumn{2}{c|}{$\rtspk = 500\gev$}\\
\hline
                  & \multicolumn{2}{c|}{$\Lum = 50~\fbinv$}
		  & \multicolumn{2}{c|}{$\Lum = 200~\fbinv$} \\
\hline
$P(e^-) / P(e^+)$    & 0/0 & {+80\%/$-$60\%}
		     & 0/0 & {+80\%/$-$60\%} \\
\hline
$\tilde{t}_1 \tilde{t}_1^*$ & 619 & 1489 
                            & 9815 & 22958  \\
\hline
$W^+W^-$ &   $11$ &   $1$ &    $<8$ &   $<1$  \\
$ZZ$     &  $< 2$ & $< 2$ &    $20$ &   $18$ \\
$W e\nu$ &   $68$ &  $20$ &  $1719$ &  $510$ \\
$e e Z$  &    $3$ &   $2$ &   $< 7$ &  $< 6$ \\ 
$q \bar{q}$, $q \neq t$ 
         &   $16$ &  $17$ &    $18$ &   $21$ \\
$t \bar{t}$ & $0$ &   $0$ &     $1$ &    $1$ \\
2-photon  &  $27$ &  $27$ &   $294$ &  $294$  \\
\hline
total background & $127$ &  $69$ & $2067$ & $851$ \\
\hline
$S / B$          & $4.9$ & $22$ & $4.7$ & $27$ \\
\hline
\end{tabular}
\caption{As in Table~\ref{tab:nev}, expected numbers of events
remaining at $\rtsth = 260\gev$ and $\rtspk = 500\gev$, with 
unpolarized and with polarized beams, after the IDA has been applied.
\label{tab:ida2}}
}
As before, 
in the channels where no event is left after the signal selection,
an upper limit corresponding to one simulated event is given in the table.

The IDA method achieves a significantly more powerful discrimination 
between signal and background than the analysis with conventional cuts.
When allowing similar background levels as for the cut-based analysis in 
Table~\ref{tab:nev}, signal efficiencies of $\effth = 0.387$ for 
$\rtsth = 260 \gev$ and $\effpk = 0.416$ for $\rtspk = 500 \gev$ 
are obtained. 

With the resulting event numbers given in Table~\ref{tab:ida2} for 
$P(e^-)/P(e^+) =$ +80\%/$-$60\%, the ratio quantity in eq.~\eqref{eq:Y} amounts to
$Y = 0.0648\pm 0.0018$ with a relative statistical error of~$2.7\%$, 
translating into
\begin{equation}
   \mst = (122.5 \pm 0.17) \gev
\label{staterrida}
\end{equation}
where the uncertainty on the mass depends on the slope $dY/dM = -0.01052$.
The higher signal efficiency and lower background achieved by the
two-step IDA results in a slightly smaller statistical uncertainty
({\it cf.} Eq.~(\ref{staterrcuts})).

\section{Experimental Systematics}
\label{sec:exp}
The high signal efficiency and low backgrounds achieved in both the
cut-based analysis (Section~\ref{sec:cut}) and the IDA
(Section~\ref{sec:ida}) deliver an excellent
statistical precision -- $\Delta\mst < 0.2\gev$.  It remains
to investigate systematic uncertainties, which were the dominant
contribution in the previous analysis of Ref.~\cite{stop}.
We considered the following important sources of systematic errors:
\label{page:listsyst}
\begin{itemize}
\item detector calibration (energy scale)
\item charm tagging
\item hadronization / fragmentation
\item neutralino mass
\item luminosity measurement
\item beam energy spectrum
\item background estimate
\end{itemize}
The first four sources pertain to the signal efficiency.
We discuss these sources in detail in the context of the sequential-cut
analysis detailed in Sec.~\ref{sec:cut} first, and then briefly report
the results obtained from the IDA method described in Sec.~\ref{sec:ida}.

\subsection{Systematics for the Sequential-Cut Analysis}
\label{sec:systcuts}
\par
Many of the kinematic quantities used in these selections depend on a
correct calibration of the calorimetry.  Based on experience from
LEP~\cite{jeten}, we assume an uncertainty of~$1\%$
on the overall energy scale, which is rather pessimistic
for a future ILC detector.  We scaled simultaneously all kinematic 
quantities through a range of $\pm 6\%$ and observed correlated
shifts in the overall selection efficiency at the two center-of-mass 
energies.  In particular, the $\pt$ cut is sensitive to this kind of
scale error, prompting us to tune the cut at~$\rtspk$ to achieve a minimal
residual uncertainty for the {\it ratio} of efficiencies, as
discussed in Section~\ref{sec:cut}.  
\par
Figure~\ref{f:scale_eff} shows how the selection efficiencies change
as a function of the scale factor.  Using our optimized $\pt$~cut shown
in Fig.~\ref{f:marg_pt}, one sees a parallel behavior at threshold
(upper solid line) and at peak (lower solid line).  This leads to
a very good cancellation for the ratio of efficiencies, as shown
by the solid line in Fig.~\ref{f:scale_reldiff}.  If we had optimized
for efficiency only, then we would have used nearly the same~$\pt$ cuts
at the peak as we use at threshold.  However, this would have given
a rather different dependence on the scale, as indicated by the
dashed line in Fig.~\ref{f:scale_eff}, and therefore a much
stronger dependence of the ratio of efficiencies on the scale,
as shown by the dashed line in Fig.~\ref{f:scale_reldiff}.
With our best cuts, an uncertainty of~$\pm 1\%$ on the calorimeter
energy scale translates into an uncertainty of less than~$0.6\%$ on
the ratio of efficiencies.
\FIGURE{
\includegraphics[width=0.65\textwidth]{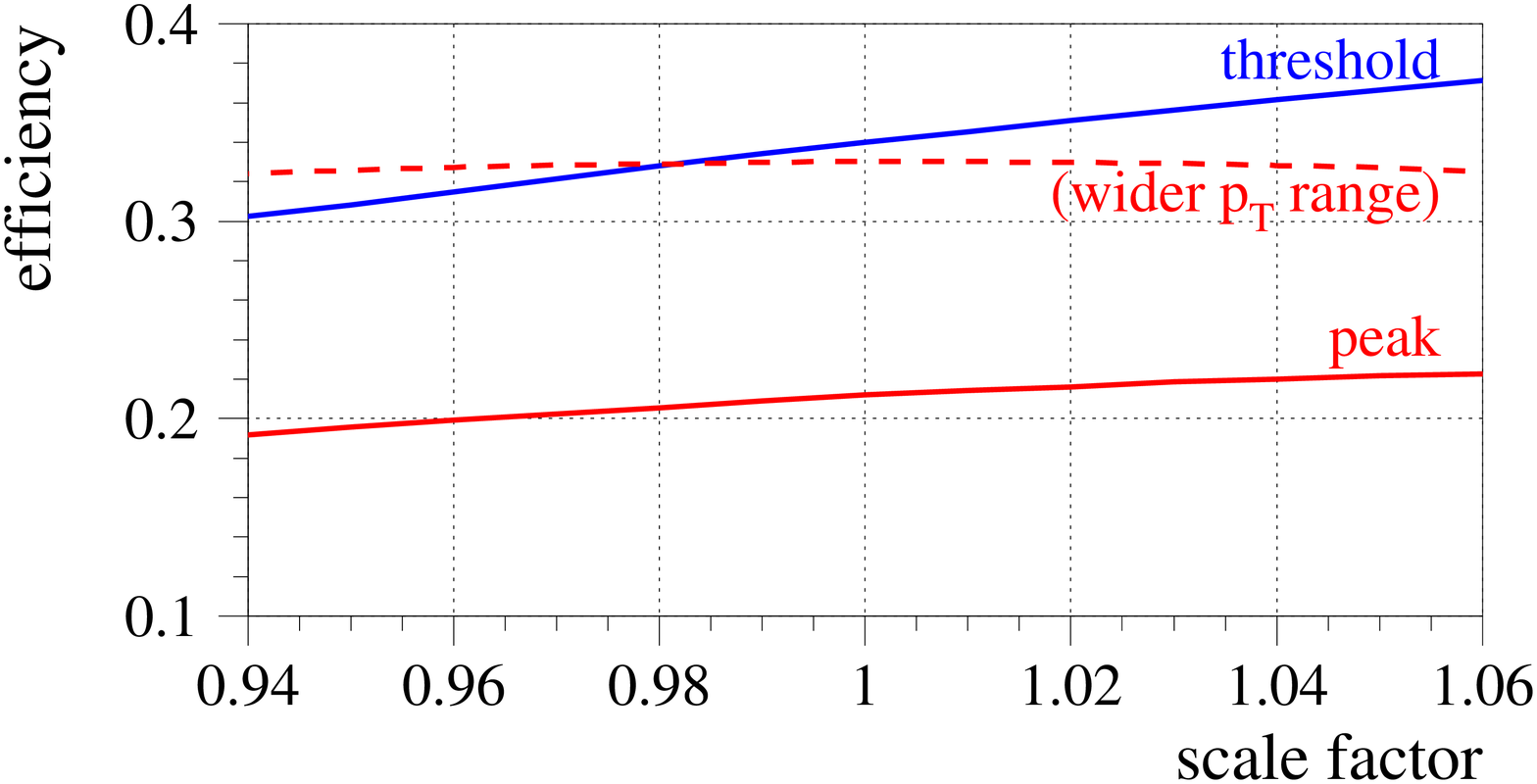}
\caption{\label{f:scale_eff} Variation of the selection efficiencies
($\effth$ and $\effpk$)
with an overall energy scale factor.  The two solid lines show the
variation obtained with our nominal cuts, at~$\rtsth$ and~$\rtspk$.
The dashed line shows what we would obtain if we applied a looser
$\pt$ cut at~$\rtspk$.}
}
\FIGURE{
\includegraphics[width=0.65\textwidth]{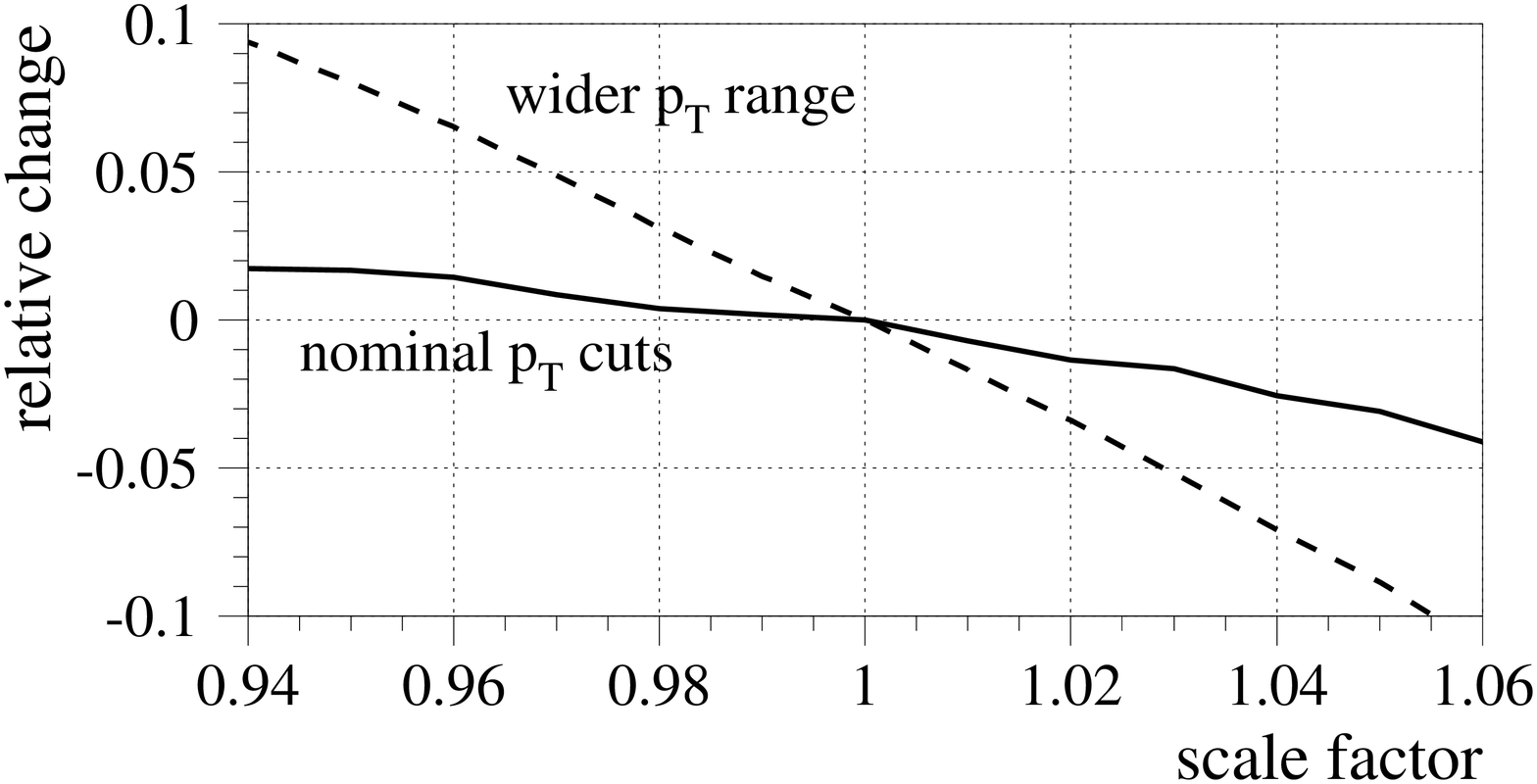}
\caption{\label{f:scale_reldiff} Relative variation of 
the ratio of efficiencies ($\effpk/\effth$)
with an overall energy scale factor.
The solid line shows a very small variation, given our
nominal cuts on~$\pt$, to be compared to a much larger
variation if we had used looser~$\pt$ cuts meant to maximize
the efficiency.}
}

\par
The efficiency for track reconstruction should be very high at
an ILC detector.  However, there is always an uncertainty in the
value for that efficiency, which we took to be about~$0.5\%$.
We propagated this uncertainty to the cut on~$\Ntracks$, since
a random loss of tracks changes the shape of the distribution
of~$\Ntracks$.  Since our cut is quite loose, however, the resulting
uncertainty on the ratio of efficiencies is negligible.
\par
Knowledge of the efficiency for charm jets for a given cut on~$\cflag$
is not easy to obtain.  Based on the work described
in Ref.~\cite{kuhlhiggs}, we assumed an uncertainty of~$0.5\%$
on the charm efficiency.  Although one might expect this uncertainty
to be correlated between the two center-of-mass energies, we
assumed no strong correlation and assign an uncertainty of~$0.5\%$
for the ratio of efficiencies.
\par
The observable $Y$ depends on the integrated luminosity at both
center-of-mass energies.  Traditionally, the luminosity is measured
using Bhabha scattering, for which highly accurate theoretical
cross-sections are available.  The limiting systematic uncertainty
for the LEP detectors comes from the acceptance of the luminosity
calorimeters.  Such an uncertainty would essentially cancel in
the ratio of luminosities.  Alternatively, one could define an
effective luminosity through another clean QED process, such
as $\epem \rightarrow \mpmm$, for which there is essentially
no theoretical or experimental systematic effect.  The precision of the
ratio of luminosities would come from the statistical uncertainty
on the number of $\mpmm$ events recorded, which we estimate
to be about $0.4\%$; this is the figure we use in this study.
\par
Apart from instrumental issues such as the energy scale, track
reconstruction efficiency, charm tagging efficiency and the
measurement of the integrated luminosity, the estimate of the 
signal efficiency will depend on the modeling of the signal itself.  
While the production of a pair of
stop quarks is well understood and can be modeled accurately,
the non-perturbative aspects of the formation of stop hadrons
which then decay into two or more jets are more 
problematic\footnote{Earlier analyses such as Ref.~\cite{stop}
neglected this important problem.}.  
We have attempted to account for this fundamental difficulty
by varying the parameter which controls the fragmentation function
in our simulations.  We used PYTHIA and the
Peterson fragmentation function, with values of the fragmentation
parameter reported by the OPAL Collaboration~\cite{fraglep}.
To be specific, we took $\epsilon_c = -0.031\pm 0.011$ and
$\epsilon_{b} = -0.0050 \pm 0.0015$, and propagated 
$\epsilon_b$ according to the assumption that 
$\epsilonstop = \epsilon_b\, (m_b/m_{\tilde{t}})^2$~\cite{fragfunc,stoplep}.
\par
We varied $\epsilon_c$ and $\epsilonstop$ independently,
and measured the efficiencies at the two center-of-mass energies.
The impact of varying $\epsilon_c$ is small.
For variations of the stop quark fragmentation, however, we find that the 
variation of the efficiencies is rather different, so the desired 
cancellation of systematic uncertainties is not achieved.  In fact, most
of the systematic uncertainty comes from the cut on~$\pt$, and has an 
{\it opposite} sense at~$\rtsth$ and~$\rtspk$.
\par
The range in $\epsilon_b$ used in our simulations is quite broad.
The more advanced measurements of $b$-quark fragmentation from 
ALEPH~\cite{alephfrag} and OPAL~\cite{opalfrag} give more constrained
values: $\epsilon_b = -0.0031\pm 0.0006$~(ALEPH) and
$\epsilon_b = -0.0041\pm 0.0004$~(OPAL), using rather different 
methodologies.  On the basis of these measurements, one could argue
that our assumed variation in $\epsilon_b$ is too large by a
factor of two. 
\par
Rather than relying on LEP measurements to predict stop quark
fragmentation, we investigated the potential of ILC data to
constrain the fragmentation.  We already noted that most of
the sensitivity to stop quark fragmentation comes from the
cut on~$\pt$; however, the change in the shape of the~$\pt$ 
distribution is small.  (The fact that the quantities chosen
for cuts are insensitive to~$\epsilonstop$ is a strong point of
the analysis, of course.)  We examined other kinematic quantities
and found a few which exhibit clear changes in shape when we
vary~$\epsilonstop$.  Four examples are shown in Fig.~\ref{f:fragplots}.
The $\Mvis/\sqrt{s}$ distribution shows pronounced shifts as a function
of~$\epsilonstop$.  Given an accumulation of a few $\times 10^4$
events at $\rtspk$, one can show that the mean of this distribution alone 
would allow a differentiation of our three values $\epsilon_b = -0.0050\pm 0.0015$
at more than ten sigma (statistical uncertainty only).  If the energy scale 
uncertainty were a problem, then one could normalize $\Mvis$ to $\Evis$ --- 
a clear distinction between the three distributions is visible near the peak
of~$\Mvis/\Evis$.  The energy of the third jet, when it exists, shows
a good sensitivity to~$\epsilonstop$.  (Recall that the jets are
ordered in decreasing energy.)  Better, perhaps, is the smaller 
of the two di-jet invariant masses formed by combining
this third jet with the first and second jets.
Although these considerations are not equivalent to a full study
of a possible measurement of the stop fragmentation, they do indicate
that a good measurement should be possible, well beyond the extrapolation of
LEP results on $\epsilon_b$ to~$\epsilonstop$ and all the attendant
assumptions behind such an extrapolation.  On this basis, we
judge that the uncertainty on the stop fragmentation would be
no larger than one-fourth of the uncertainty obtained by comparing
simulations with $\epsilon_b = -0.0035$,~$-0.0050$ and~$-0.0065$,
which corresponds to 
$\Delta\epsilonstop = (2.5\times 10^{-6})/4 = 0.6\times 10^{-6}$.

\FIGURE{
\mbox{\includegraphics[width=0.45\textwidth]{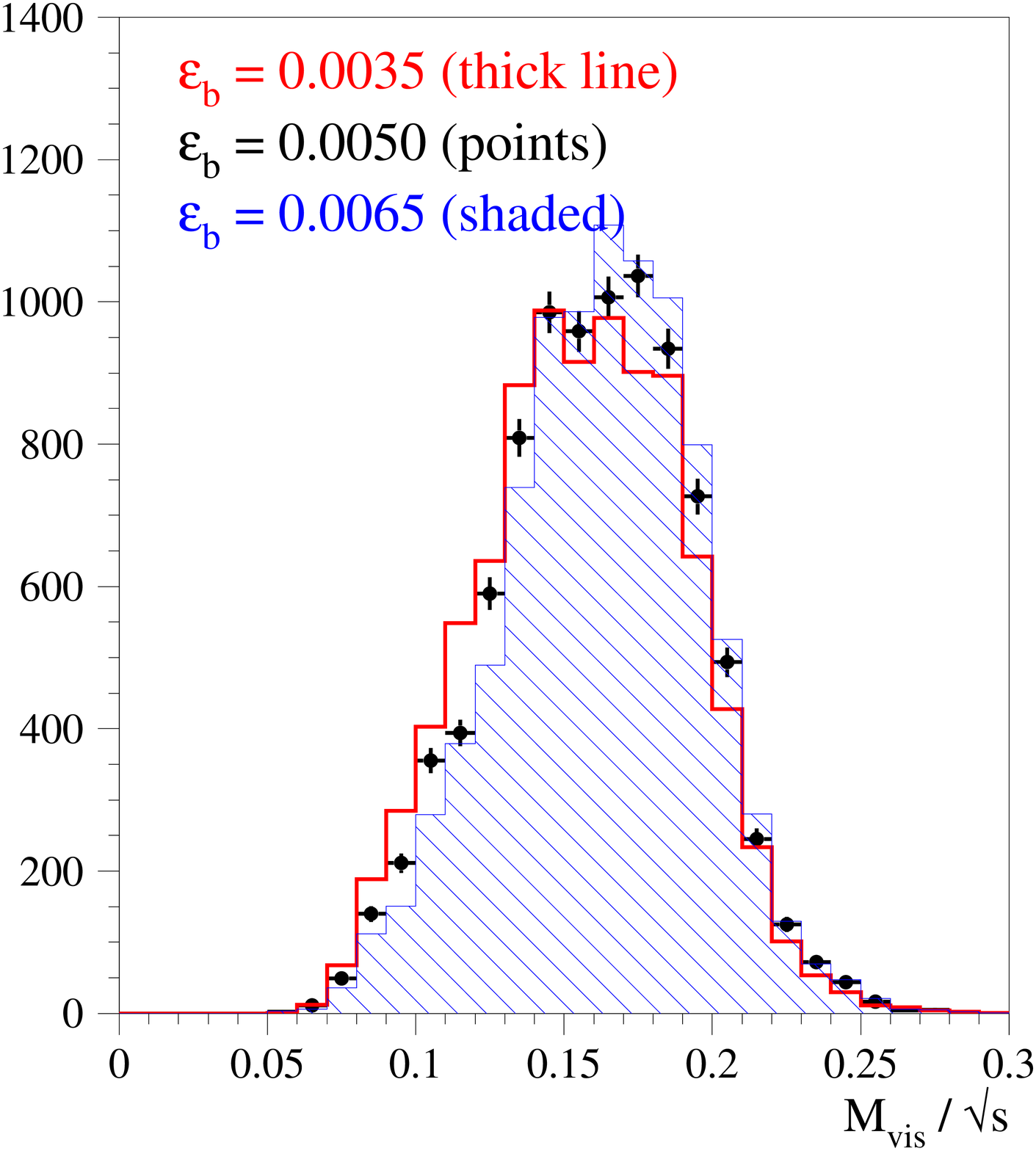}
      \includegraphics[width=0.45\textwidth]{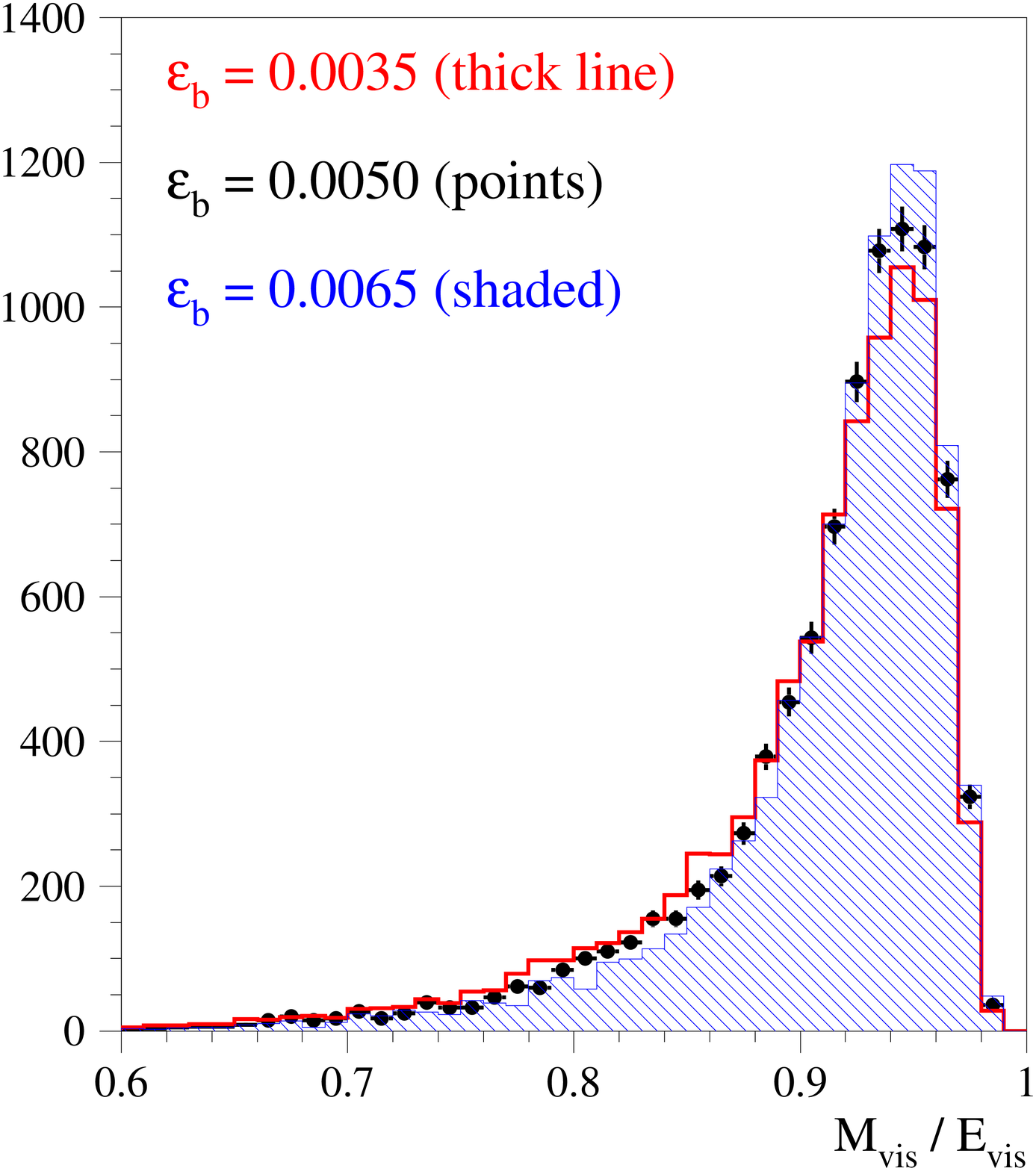}}
\mbox{\includegraphics[width=0.45\textwidth]{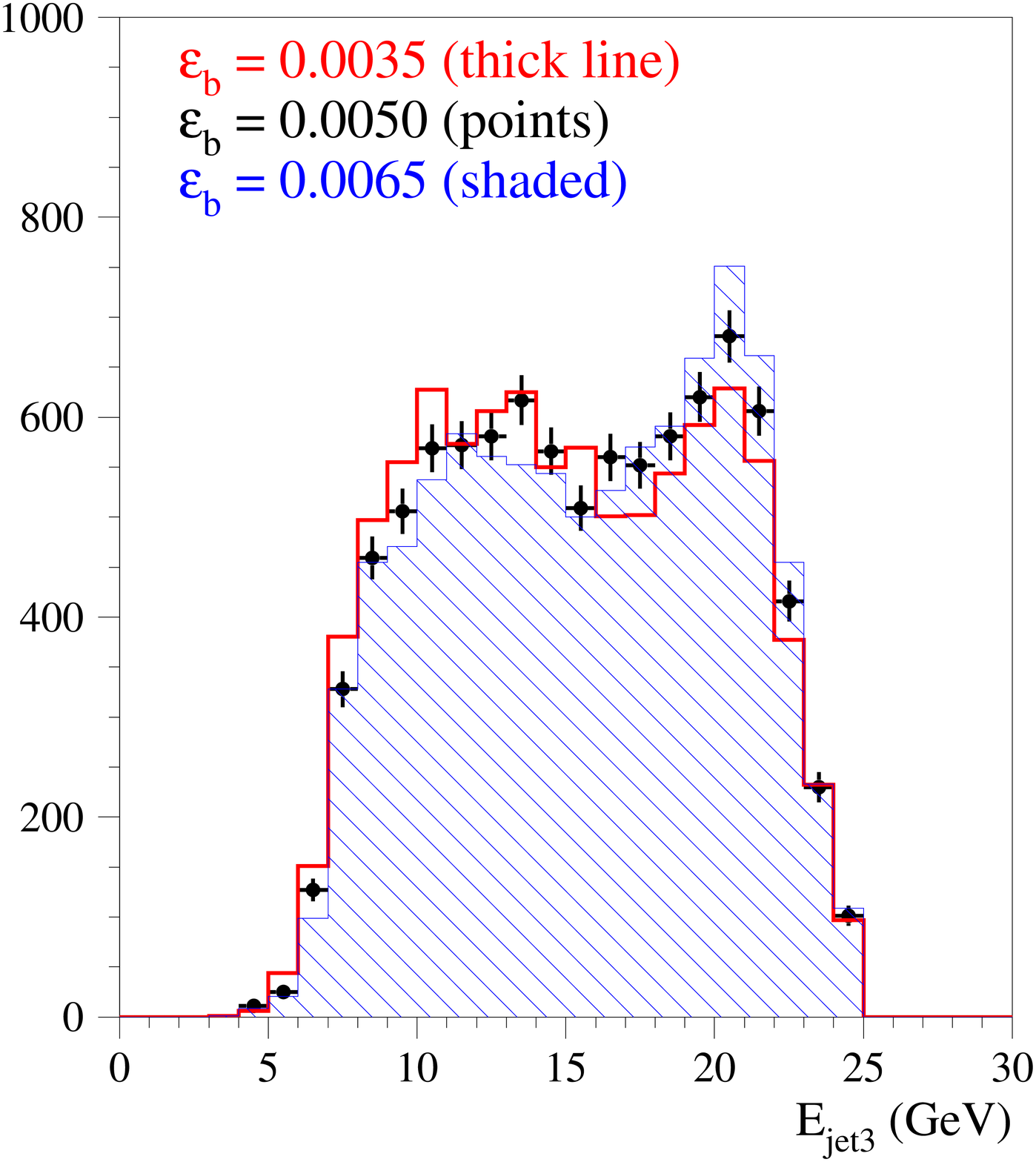}
      \includegraphics[width=0.45\textwidth]{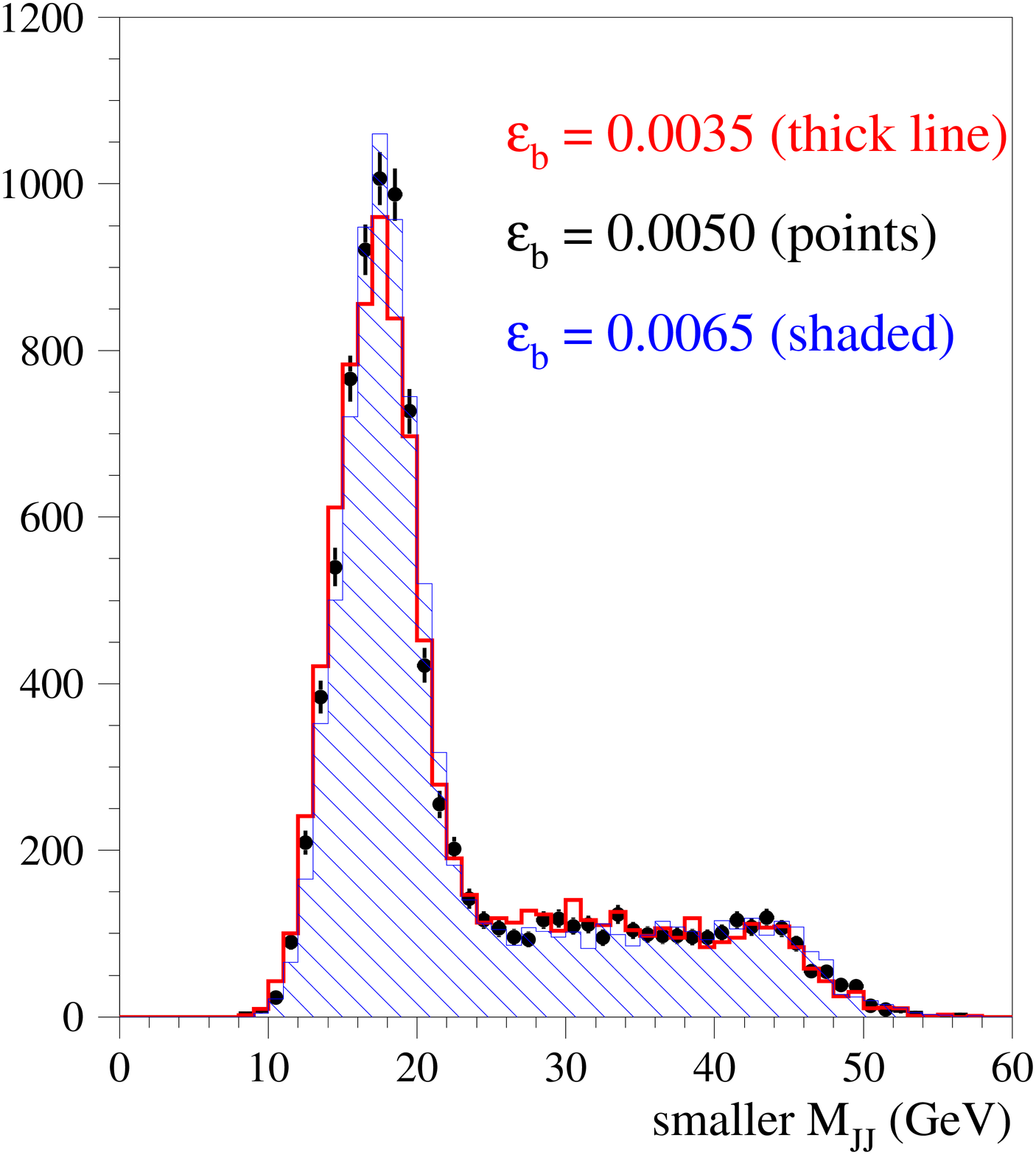}}
\caption{\label{f:fragplots} 
Changes in kinematic distributions at $\rtspk$ for different assumed
values of~$\epsilonstop$, which are related directly to the listed
values of~$\epsilon_b$ through $\epsilonstop = \epsilon_b\, (m_b/m_{\tilde{t}})^2$.
The solid points with error bars show the 
distribution with $\epsilon_b = -0.0050$, our default choice.
The thick, unshaded red histogram shows $\epsilon_b = -0.0035$,
and the thin, shaded blue histogram shows $\epsilon_b = -0.0065$.
}
}

\par
Another empirical quantity which induces an uncertainty on the selection
efficiency is the mass of the neutralino, $\mneu1$.  The mass difference
$\mst-\mneu1$ clearly impacts the kinematic distributions, so the efficiency
estimated from the simulation depends directly --- and dramatically ---
on~$\mneu1$.  We simulated a sample with $\mneu1 = 108.2\gev$, which is
one$\gev$ higher than our default value.  The relative change in the selection
efficiencies is roughly $10\%$.  Since the changes are parallel, the ratio of
efficiencies change by only $2.8\%$, once again illustrating the robustness of
this method.  Other studies have shown~\cite{heavyq} that $\mneu1$ can be
measured with an accuracy of $0.3\gev$ or better,  so we assign an uncertainty
of $0.8\%$ due to the unknown neutralino mass.
\par
The predicted cross-sections depend on the beam energy and the beam
energy spectrum.  Due to beamstrahlung and other effects, the
mean energy can be significantly lower than the peak value.
While we used CIRCE for taking this fact into account, the question
remains how well a program such as CIRCE can be validated  using
real data.  This question has been addressed by several 
authors, using, for example, Bhabha scattering and radiative returns 
to the $Z$~pole~\cite{beamspectrum}.  The studies
indicate that models for the spectrum and the beam energy can be 
constrained directly from the data to an accuracy on the order
of $0.1\gev$.  We include this uncertainty as a direct uncertainty
on $\mst$, but not on the observable~$Y$.
\par
Finally, we must consider uncertainties on the estimated contributions
from background processes.  The SM backgrounds fall naturally in two
categories:  two-photon interactions, which are difficult to predict,
and the others, which involve high-$\pt$ electro-weak processes, for
which direct theoretical calculations are reliable.  We also 
consider background contributions from the production of other 
supersymmetric particles.
\par
Two-photon interactions cannot be fully described by perturbative QCD,
and so phenomenological models are required~\cite{twophoton_review}.
These must be tuned to match real data, which is difficult due to the fact
that most two-photon scattering events emit particles that are lost
down the beam pipe.  Parameters pertaining to the softest interactions 
are difficult to constrain; fortunately, such interactions are easily
eliminated by our cuts on~$\pt$, $\Ntracks$ and~$\Evis$.  Many of the
events coming in at higher~$\pt$ can be described using models with
a basis in perturbative QCD.  The investigations of the photon structure functions
by the LEP Collaborations illustrate the procedure of tuning parameters
and confronting the models with real data, leading to interesting conclusions 
about the success of the various 
models~\cite{twophoton_LEP,twophoton_OPAL,twophoton_LEPWG}.
It is not straightforward to translate those conclusions into
constraints on our two-photon background, although Figs.~19,~21 and~23
in the report from the LEP Working Group~\cite{twophoton_LEPWG}
and Figs.~5-7 in the OPAL paper~\cite{twophoton_OPAL} are quite relevant 
for our study, and indicate that modeling the tails of the $\pt$
distribution at the 20\% level should be possible.  Assuming that
the study of two-photon interactions would be greatly extended at the ILC, 
we assign a 20\% uncertainty to the background estimate for two-photon interactions.
The resulting relative uncertainty on the $Y$ observable is~$0.8\%$.
\par
The dominant background is $\epem\rightarrow We\nu$, according to Table~\ref{tab:nev}
(and Table~\ref{tab:ida2}).
A precise prediction of this background requires accurate measurements 
of this process combined with the calculation of
higher-order radiative corrections. While a complete NLO calculation of that
process is missing, a recent result for the related process of $W$ pair
production~\cite{wwnlo} suggests that a NLO calculation of $W e \nu$ is
feasible within the next years with an error remaining well below~$0.5\%$. 
The impact on~$Y$ is negligible, on the order of $0.1\%$, relative.
\par
A summary of the experimental systematic uncertainties for the
sequential-cut analysis is shown in Table~\ref{tab:sys}. 
A good cancellation of experimental systematics is obtained,
except for the stop quark fragmentation uncertainty
and the background estimation.
The goal of the new method is therefore fairly well achieved
with this set of sequential cuts.
The implications for the measurement of the observable~$Y$ and
the inferred mass~$\mst$ are discussed in Sec.~\ref{sec:comb}.

\TABLE{
\begin{tabular}{|l|l|c|c|c|}
\hline
 & error on &  \multicolumn{2}{c|}{relative shift in expected signal yield (\%)} & error on\\
variable & variable & $\rtsth = 260 \gev$ & $\rtspk = 500 \gev$ & $Y$ (\%) \\
\hline
energy scale             & $1\%$ &  $3.7$ &  $3.1$ & $0.6$ 
\\
tracking efficiency      & $0.5\%$ &  \multicolumn{3}{c|}{negligible}
\\
charm tagging efficiency & $0.5\%$ &  \multicolumn{3}{c|}{taken to be $0.5$}
\\
luminosity               & - & $0.4$ & $0.2$ & $0.4$ 
\\
charm fragmentation      & $0.011$ & $0.3$ & $0.8$ & $0.6$
\\
stop fragmentation       & $0.6\times 10^{-6}$ & $0.6$ & $0.2$  & $0.7$ 
\\
neutralino mass          & $0.3\gev$ & $3.8$ & $3.0$ & $0.8$
\\
background estimate      & - & $0.8$ & $0.1$ & $0.8$
\\
\hline
\end{tabular}
\caption{\label{tab:sys}
Evaluation of experimental uncertainties on~$Y$,
for the sequential-cut analysis.
The last column gives the relative uncertainty on~$Y$.}
}

\subsection{Systematics for the Iterative Discriminant Analysis}
\label{sec:systida}
\par
We evaluated the impact of the sources of systematics listed
on page~\pageref{page:listsyst} in a manner similar to the
methods of Sec.~\ref{sec:systcuts}.  We scaled all kinematic
inputs to the IDA according to an overall energy scale uncertainty.
The systematic uncertainty from the number of tracks is
assumed to be negligible.  The variations in the charm and stop
quark fragmentation functions were evaluated as before.
The sensitivity to~$\mneu1$ and the uncertainty on the background
estimate were evaluated precisely as above.
The luminosity uncertainty is, of course, the same as
in the sequential-cut analysis.
\par
The resulting systematic uncertainties are listed in
Table~\ref{tab:sysida}.  We observe a much larger uncertainty coming
from the scale uncertainty as compared to the sequential-cut analysis
({\it cf.} Table~\ref{tab:sys}).  With multi-variate methods such as
the IDA, it is difficult to ascertain what role any given quantity
plays in the final output variable, so no dissection of the IDA to
reveal the sensitivities to the energy scale is possible.
Furthermore, one cannot tune the operation of the IDA in order
to balance efficiencies for each quantity, as we did for
thrust~$T$ and event-$\pt$ in the sequential-cut analysis.
For this kind of precision measurement, it would appear that the
better discrimination of signal and background provided by the IDA
is of limited value in light of the larger and uncontrollable 
sensitivity to experimental sources of systematic uncertainty.
Nonetheless, when performing a measurement with real data, one
would welcome an alternative analysis in order to check the
robustness and stability of the measurement.
\par
In Section~\ref{S:discovery}, we show the power of the~IDA
in the discovery of a light stop quark.

\TABLE{
\begin{tabular}{|l|l|l|l|l|}
\hline
 & error on &  \multicolumn{2}{c|}{relative shift in expected signal yield (\%)} & error on\\
variable & variable & $\rtsth = 260 \gev$ & $\rtspk = 500 \gev$ & $Y$ (\%) \\
\hline
energy scale              & $1\%$ &  $3.4$ &  $1.3$ & $2.3$ 
\\
tracking efficiency       & $0.5\%$ &  \multicolumn{3}{c|}{negligible}
\\
charm tagging efficiency  & $0.5\%$ &  \multicolumn{3}{c|}{taken to be $0.5$}
\\
luminosity                & - & $0.4$ & $0.2$ & $0.4$ 
\\
charm fragmentation       & $0.011$ & $0.1$ & $0.6$ & $0.5$
\\
stop fragmentation        & $0.6\times 10^{-6}$ & $0.1$ & $0.8$  & $0.7$ 
\\
neutralino mass           & $0.3\gev$ & $3.7$ & $1.6$ & $2.2$
\\
background estimate       & - & $0.3$ & $0.2$ & $0.1$
\\
\hline
\end{tabular}
\caption{\label{tab:sysida}
Evaluation of experimental uncertainties on~$Y$,
for the~IDA.  The last column gives the relative uncertainty on~$Y$.}
}

\section{Theoretical Uncertainties}
\label{sec:th}

The inference of the stop mass from stop cross-section measurements
requires precise theoretical calculations for the cross-sections. The stop
production cross-section receives large corrections in particular from QCD
gluon exchange between the final state stops. Near threshold, when the stop
quarks are slowly moving, these effects become very large, which is the
well-known Coulomb correction \cite{coulomb}. The NLO
QCD corrections to stop production have been computed several years ago
\cite{calvin} and it was found that the corrections range between about 10\% at
high energies and up to 100\% near threshold. This shows that higher-order
corrections are crucial.

Over the last few years, sophisticated techniques have been developed for 
calculating top-pair production at NNLO~\cite{topthr}.
Near threshold, they include resummation of terms of order
$O(\alpha_{\rm s}/v)$ for the low velocity $v$ of the top-quarks. For the
production of scalar quarks, similar calculations are not yet done. However,
one can use partial results to obtain a reliable estimate of the
uncertainty of the NNLO corrections. At NNLO order, several contributions enter
in the computation. The largest effect near threshold arises from the
Coulomb correction. The impact of the Coulomb corrections is calculated
through NNLO order \cite{clnnlo}, including resummation via non-relativistic
QCD. Technically, here the non-relativistic Schr\"odinger equation is used for
computing the Coulomb effects \cite{schreq}.

Similar to the case of top pair production, it is found that the NNLO term to stop
pair production is of similar order of magnitude as the NLO term, {\it i.e.}, the
perturbation series is converging rather slowly. From the behavior of the
perturbation series and the remaining scale dependence, the size of the missing
higher-order contributions is estimated to be around 7\% at 260 GeV and 2.5\% at
500 GeV. 

However, we want to point out that several improvements to this straightforward
approach could be made. Besides the large Coulomb-type corrections of order
$O(\alpha_{\rm s}/v)$, there are also potentially large logarithmic
contributions $O(\log(\alpha_{\rm s}/v))$. They can be resummed with more
sophisticated techniques, for instance velocity non-relativistic QCD~\cite{vnrqcd}. 
Using the results of Ref.~\cite{vnrqcd:scalar} for the NLO
corrections to squark pair production, it is found that the uncertainty with
respect to the NLO computation is reduced significantly. A similar
improvement can be expected at the NNLO level. In addition, instead of directly
computing the total cross-section near threshold, one can describe it through
moments~\cite{moments} that avoid the non-perturbative contribution of
stopponium bound states that can form just below the nominal stop-pair
threshold. With these refinements it is expected that the theoretical
uncertainty can be brought down by a factor of two (however the actual
calculation remains for the future). So here an uncertainty of 3.5\% at 260 GeV
and 1\% at 500 GeV are assumed.

Besides the QCD corrections, the electroweak corrections need to be considered.
The NLO electroweak corrections have been computed  \cite{ewnlo}, and found to
amount to several per-cent. While they need to be taken
into account, the NNLO corrections are expected to be much less than 1\%, with
the exception of leading initial- and final-state QED corrections that can
easily be resummed to higher orders.

Combining the two errors, a total theoretical error of 4\% at $260\gev$ and 1.5\%
at $500\gev$ can be assigned.  Pessimistically, we add these two uncertainties
linearly, and assign a theoretical uncertainty of $5.5\%$ for the quantity $Y$.


\section{Results and Implications}
\label{sec:comb}
\par
We derive the expected measurement error on the stop quark mass
and discuss the implications for particle physics predictions
of the relic density of dark matter.
We also discuss the luminosity needed to discover a light
stop quark in this scenario, using the IDA method.

\subsection{Precision on the Stop Quark Mass}
\par
A final assessment of the achievable precision on the stop mass
will be based on the statistical and all systematic uncertainties.
Table~\ref{tab:sum} summarizes these uncertainties for the
observable~$Y$ defined by Eq.~\eqref{eq:Y}.
One sees that the IDA method achieves a smaller statistical
uncertainty on~$Y$ at the cost of a larger experimental
systematic uncertainty.  It would be important, in a measurement
with real data, to implement two methods as we have done here,
and check the consistency of the results.
\par
The stop quark mass is inferred from the measured values of
the observable~$Y$ following the example described in 
Sec.~\ref{sec:method}.  The differing efficiencies for
the sequential-cut and IDA methods lead to different central
values for~$Y$ and for the slope $dY/dM$ at that point.
The inferred uncertainties on the stop quark mass are
summarized in Table~\ref{tab:mstoperr} and are similar
for the two analyses.
We conclude that the stop quark mass could be measured
with an uncertainty of $\Delta\mst = 0.42\gev$.
\par
We investigated the dependence of the measurement error on the
integrated luminosity.  There is very little change in the 
statistical uncertainty if we increase the luminosity on peak,
but the variation with the luminosity at $\rtsth = 260\gev$
is interesting -- see Fig.~\ref{F:errorlum}.  The experimental
uncertainty is dominated by the statistical contribution,
so a decrease in the luminosity from our assumed value of
$\Lumth = 50~\fbinv$ has a significant impact.  On the
other hand, the theoretical uncertainty is very large by
comparison, so increasing $\Lumth$ hardly improves the
total error on~$\mst$.  A luminosity in the range
$30~\fbinv < \Lumth < 80~\fbinv$ would appear to be optimal,
for this analysis.
\par
The dominant uncertainty comes from the theoretical calculation
of the signal cross-section.  As discussed in Sec.~\ref{sec:th},
this uncertainty comes mainly from higher-order corrections which
are not easily summed at threshold.  The estimate of this theory 
error relies on present computational techniques and some expectations 
on how they might improve in the future. However, the progress in calculations 
of radiative corrections can not really be predicted, so the assumed value for the
theoretical uncertainty at the time when ILC is running might well be somewhat
different than the value reported in Table~\ref{tab:sum}. In particular,
history has shown that people working on loop
computations often overcame big problems with unexpected ingenuity, in order to
be able to make most of precise measurements. Therefore, in the following, the
combined error in Table~\ref{tab:sum} will be taken as a conservative estimate.
If one were to set aside the theoretical error on the cross section, then
the total experimental error is quite small, amounting to $3.5\%$--$4.3\%$
on~$Y$.  In this case, the error on the stop quark mass would be a little
larger than $\Delta\mst = 0.2\gev$.

\TABLE{
\begin{tabular}{|l|c|c|}
\hline
error source for $Y$ & sequential cuts & IDA method \\
\hline
detector effects                      & 0.9 &   2.4 \\
charm  fragmentation                  & 0.6 &   0.5 \\
stop fragmentation                    & 0.7 &   0.7 \\
neutralino mass                       & 0.8 &   2.2 \\
background contribution               & 0.8 &   0.1 \\
sum of experimental systematics       & 1.7 &   3.4 \\
\hline
statistical                           & 3.1 &   2.7 \\
sum of experimental errors            & 3.5 &   4.3 \\
\hline
theory for signal cross-section       & 5.5 &   5.5 \\
\hline
total error $\Delta Y$                & 6.5 &   7.0 \\
\hline
\end{tabular}
\caption{Summary of relative statistical and systematic uncertainties 
(in percent) on the observable~$Y$.
\label{tab:sum}}
}

\TABLE{
\begin{tabular}{|l|c|c|}
\hline
 & \multicolumn{2}{c|}{measurement error $\Delta\mst$ (GeV)} \\
error category & sequential cuts & IDA method \\
\hline
statistical                             & $0.19$   & $0.17$ \\
sum of experimental systematics on $Y$  & $0.10$   & $0.21$ \\
beam spectrum and calibration           & $0.1~$   & $0.1~$ \\
sum of experimental errors              & $0.24$   & $0.28$ \\
sum of all experimental and theoretical errors
                                        & $0.42$   & $0.44$ \\
\hline
\end{tabular}
\caption{Estimated measurement errors (in~GeV) on the stop quark mass
\label{tab:mstoperr}}
}

\FIGURE{
\includegraphics[width=0.65\textwidth]{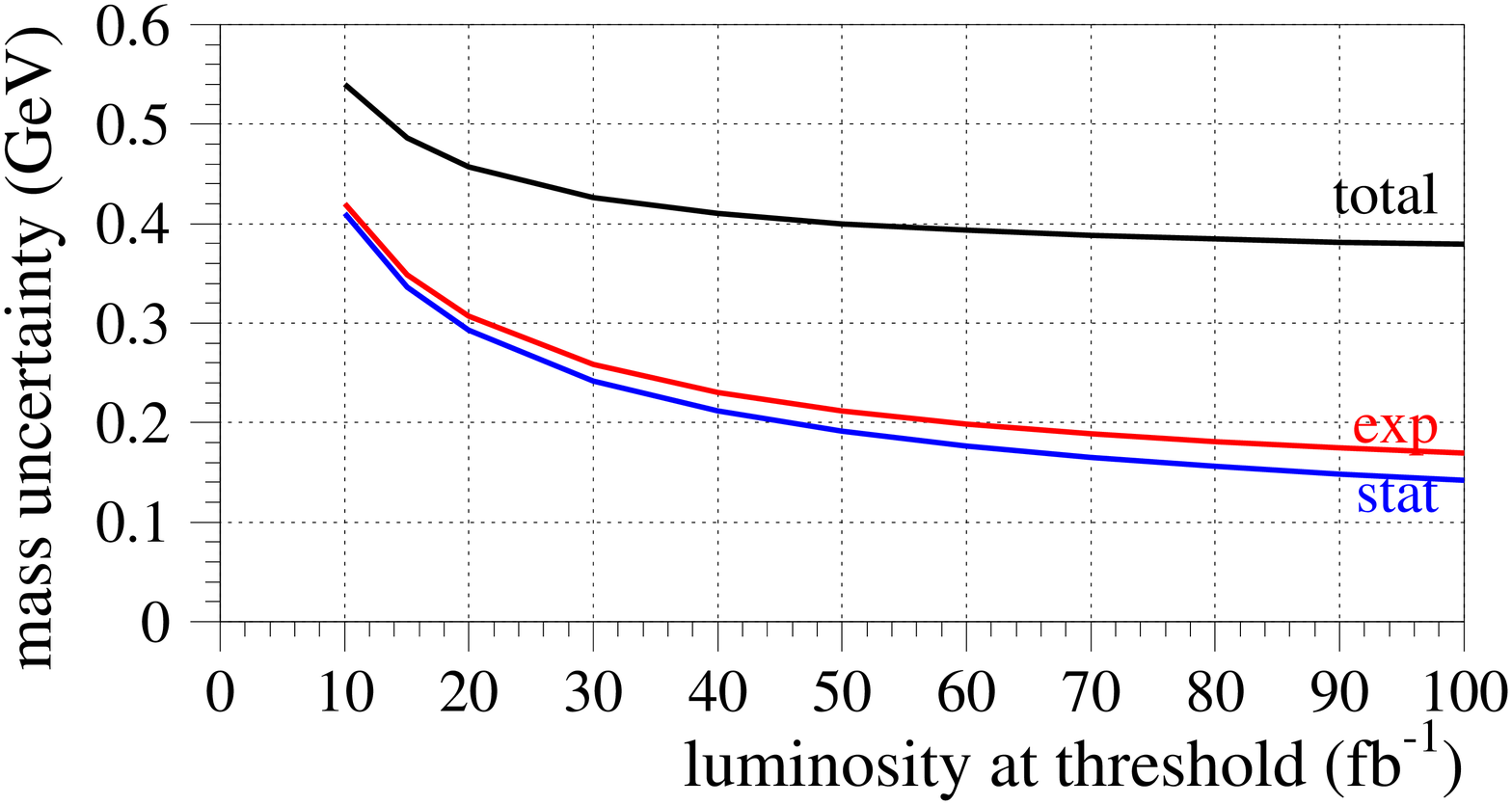}
\caption{\label{F:errorlum} 
Decrease of the statistical uncertainty (blue line),
total experimental uncertainty (red line) and total
uncertainty on~$\mst$ (black line), as a function of
the integrated luminosity~$\Lumth$ at~$\rtsth$.
}
}

\subsection{Comparison with Previous Results}
\par
A previous study investigated the potential of the ILC
running at $\sqrt{s} = 500$~GeV to discover a light stop
quark and measure its parameters~\cite{stop}.  It was assumed
that $250~\fbinv$ would be taken at two beam polarization
combinations: $P(e^-)/P(e^+) = +80\%/-60\%$ and $-80\%/+60\%$.
Measurements of the stop squark production cross sections
at these two polarizations are sufficient to deduce the
mixing angle and mass of the stop squark.  A host of systematic
uncertainties was considered, with the conclusions that
the {\it absolute} cross-sections could be measured to 
$1.3\%$--$2.4\%$, dominated by experimental systematic
uncertainties (the statistical uncertainty was~$0.8\%$).
Under the given theoretical scenario,
the total error on the stop quark mass was 
estimated\footnote{Note that the error of 1.2~GeV is slightly larger than reported in
Eq.~(17) in Ref.~\cite{stop}, since we are using the scenario from
Ref.~\cite{heavyq} with large slepton masses, leading to a larger neutralino
mass error, which in turn increases the stop mass uncertainty.}
to be~$\Delta\mst = 1.2$~GeV.

Our theoretical scenario coincides with the one studied in 
Ref.~\cite{stop}, and the method proposed here leads to
a total error on the stop quark mass that is more than
two times smaller: $\Delta\mst = 0.42$~GeV,
even though a much smaller integrated luminosity is assumed.
This improvement is certainly valuable, and is quite helpful
for the calculation of the relic density, as we discuss below.
We would like to point out, however, that the basis for the
experimental analysis in Ref.~\cite{stop} differs significantly
from what was used for the present analysis.  In particular,
the fragmentation of the stop squark and of the charm quark
produced in its decay was not simulated in Ref.~\cite{stop},
leading to very different signal characteristics which are
not realistic.  For example, the number of jets was almost
always two, which contrasts starkly with the present study in
which typically one or two additional jets are found due to
the process of fragmentation.  Thus the requirement in 
Ref.~\cite{stop} of exactly two jets, which is very effective
at suppressing the simulated $W e\nu$ background, leads in
practice to a very low efficiency and large sensitivity to
the modeling of the fragmentation practice.  Ref.~\cite{stop}
assumed a~$1\%$ systematic uncertainty on the {\it absolute}
efficiency due to fragmentation, which dominates the
total uncertainty on the cross-section measurement.
As we have seen in the present study, this is likely to be
significantly underestimated.   Furthermore, no serious
assessment of the theoretical uncertainty on the cross-section
was given in Ref.~\cite{stop}; this uncertainty was assumed
to be negligible.  As we discuss in Section~\ref{sec:th} above,
this assumption is not justified, and with existing techniques
we estimate the theoretical uncertainty to be as large as~$2.5\%$
at $\sqrt{s} = 500$~GeV, which we assume will be reduced 
to~$\approx 1\%$ by the time the ILC is running.
\par
We evaluated the efficiency of the selection in Ref.~\cite{stop}
using the signal samples generated for the present study.
We find that the requirement $\Njets = 2$ delivers an efficiency
of only~$7.3\%$, to be compared to $\approx 18\%$ reported in~\cite{stop}.
This would increase the statistical uncertainty on the cross-section
measurement from~$0.8\%$ to~$1.2\%$.
If we relax the cut to $\Njets \ge 2$, the efficiency climbs to~$21\%$,
though clearly the background would become too large with
this cut.  We evaluated the scale dependence and found it to
be negligible, since there are relatively few energy-based cuts
in~\cite{stop}.  We evaluated the systematics for the
stop and charm fragmentation, and found a very large sensitivity
to the fragmentation, on the order of~$5\%$, due to the requirement
that $\Njets = 2$ only.
\par
We conclude that the uncertainty on the stop quark mass, $\Dmst = 1.2\gev$,
reported in Ref.~\cite{stop} was underestimated,
so that our present result $\Delta\mst = 0.42$~GeV represents
a major step forward.

\subsection{Implications for Relic Density Calculation}
\par
Precise measurements of supersymmetric particle properties at the LHC and ILC
can be used to compute the dark matter relic abundance so as to compare with
cosmological observations. If stop-neutralino co-annihilation is relevant, as
in the scenario studied here \cite{heavyq}, it is important to measure the
stop-neutralino mass difference very precisely.  The extraction of the
neutralino properties, in particular the lightest neutralino mass, is studied
in detail in Ref.~\cite{heavyq}. It is found that a high precision of
$\Delta\mneu{1} \approx 0.3$ GeV for the lightest neutralino mass can be
achieved at the ILC, and also the other neutralino parameter can be inferred
rather well.

The limiting factor in the accuracy of the dark matter estimation is therefore
the precision of the measurement of the scalar top quark mass. As discussed in the previous
section, an older study using cross-section measurements at $\sqrt{s} =500$~GeV 
found $\Dmst = 1.2$ GeV and $|\cos\theta_{\tilde{t}}| < 0.077$
for the stop mass and mixing angle~\cite{stop,heavyq}.
Based on these expected experimental results, the relic dark matter density is
computed with the codes described in Ref.~\cite{Balazs:2004bu,morr}.
Fig.~\ref{fig:dm} shows the result of a scan over the MSSM parameter space. 
The scattered gray dots indicate the region allowed by the collider
experimental uncertainty, as a function of the measured stop mass. The
horizontal bands depict the relic density as measured by WMAP~\cite{wmap} with
one and two standard deviation errors. 
Here, $\Omega_{\rm CDM}$ is the ratio of the dark matter
energy density to the critical density $\rho_c = 2H_0^2/(8\pi G_{\rm N})$,
with the Hubble constant $H_0 = h\times 100$~km/s/Mpc and 
Newton's constant $G_{\rm N}$.
At the 1$\sigma$ level, the astrophysical
observations lead to $0.103 < \Omega_{\rm CDM} h^2 < 0.116$. With a stop mass
measurement error of $m_{\tilde{t}_1} = (122.5 \pm 1.2) \gev$, the relic
density can be predicted to $0.082 < \Omega_{\rm CDM} h^2 < 0.139$ at the
1$\sigma$ level.
With the new result of this work, $\Delta\mst = 0.42\gev$,
the relic density can be computed much more precisely, yielding
the result $0.096 < \Omega_{\rm CDM} h^2 < 0.124$. This precision is very
comparable to the direct WMAP measurement, as indicated by the black dots in
Fig.~\ref{fig:dm}.

\FIGURE{
\epsfig{file=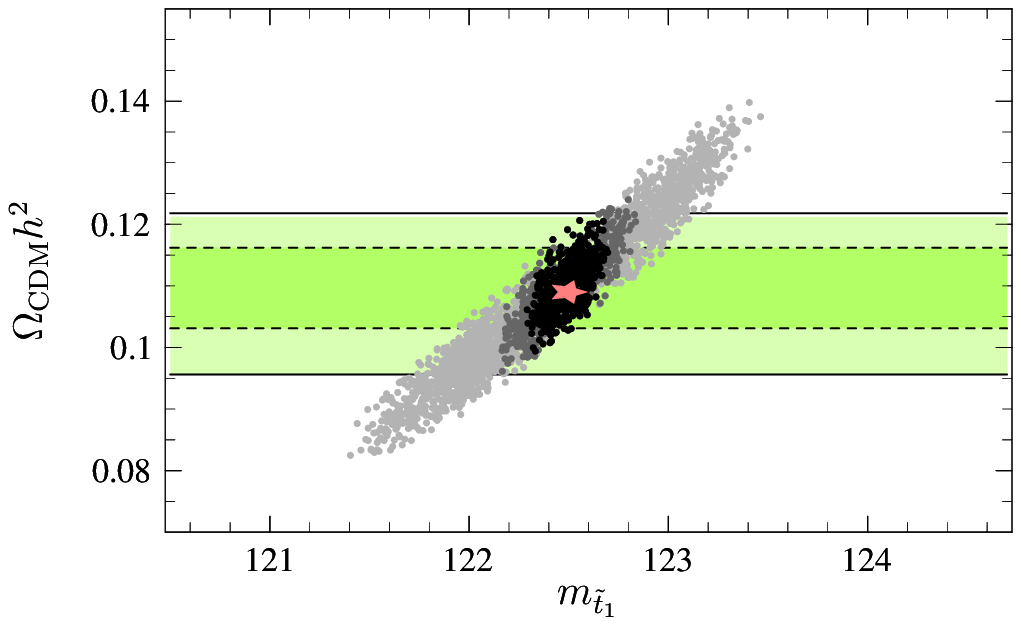, width=15cm}
\caption{Computation of dark matter relic abundance $\Omega_{\rm CDM} h^2$
taking into account estimated experimental errors for stop, chargino, neutralino
and Higgs sector measurements at future colliders. The dots correspond to
a scan over the 1$\sigma$ ($\Delta \chi^2 \leq 1$) region allowed by the 
experimental errors, as a
function of the measured stop mass, for a mass measurement error of 1.2 GeV
(light gray dots), 0.42 GeV (dark gray dots) and 0.24 GeV (black dots). The underlying scenario used as input is indicated by the red (light) star. The horizontal shaded bands show the 1$\sigma$ and 2$\sigma$ constraints on the relic
density measured by WMAP.\label{fig:dm}}
}

As pointed out above, the estimate of a stop mass error of $\Dmst = 0.42\gev$ 
is based on a rather conservative evaluation of systematic errors.
In particular, this value is dominated by the conjectured theory error on the
prediction of signal and background cross-sections. If on the other hand, with
progress in calculation methods, the theory error could be reduced to a
sub-dominant level, the remaining statistical and systematic experimental errors
would give a stop mass error of $\Dmst = 0.24\gev$ for the cut-based analysis and  
$\Dmst = 0.27\gev$ for the IDA.
The amelioration of the prediction for the dark matter relic density due to
this improvement in stop mass precision is illustrated in Fig.~\ref{fig:dm}.

For this accuracy of the stop mass measurement, the uncertainty of
the dark matter prediction becomes limited due to the expected experimental
errors in the lightest neutralino mass and mixing angles, which we have taken
from Ref.~\cite{heavyq}. 
As a result, taking an error of $\Dmst = 0.24\gev$ for the stop mass, we find 
$0.099 < \Omega_{\rm CDM} h^2 < 0.121$, which is only a small improvement in the
precision of the dark matter density prediction with respect to $\Dmst = 0.42 \gev$.

\subsection{Discovery of the Light Stop Quark}
\label{S:discovery}
\par
The main focus of this paper is the measurement of the stop quark mass.
It is interesting, nonetheless, to consider the utility of these
selections for discovering the light stop quark at the ILC. 
The IDA-based selection, in particular, achieves a very low background
and a high efficiency -- see Table~\ref{tab:ida2} in Subsection~\ref{sec:ida}.
\par
We examined this issue assuming that the ILC collects data at $\rtspk = 500$~GeV, 
with {\it unpolarized} beams, as one might expect at start-up.  The signal 
cross-section for this scenario is $\sigma_{\tilde{t}} = 118$~fb.  The nominal
IDA selection efficiency is $\epsilon = 0.416$ and the background cross-section 
for unpolarized beams is $\sigma_b = 10.3$~fb.  Tightening the selection to reduce
the background improves the sensitivity of the analysis only very slightly.
This information allows a calculation of the expected tail probability or
$p$-value as a function of integrated luminosity,~$\Lum$.  Specifically, we computed
the $p$-value setting the hypothetical number of observed events equal to the mean of the
corresponding Poisson distributions (signal and background), as a function of~$\Lum$.
The result is shown in Fig.~\ref{f:pvalues} by the thick red line.
The black dots on the line show hypothetical integral numbers of observed events,
starting at $N = 1$ for $\Lum = 0.02~\fbinv$.  The plot clearly indicates
that a luminosity of only $\Lum\approx 240~\pbinv$ would produce 
eleven observed events, on average, and the significance of ten signal events 
over the expected background would be more than~$5\sigma$.
The uncertainty on the background estimate and the signal efficiency have
a negligible impact on this result.
\par
This example applies only to our given scenario, with $\mst = 122.5\gev$,
$\mneu1 = 107.2\gev$ and $\cos\theta_{\tilde{t}} = 0.01$.  Further investigations
would be needed in order to understand how well this IDA selection would perform for 
other mass and mixing combinations.

\FIGURE{
\includegraphics[width=0.65\textwidth]{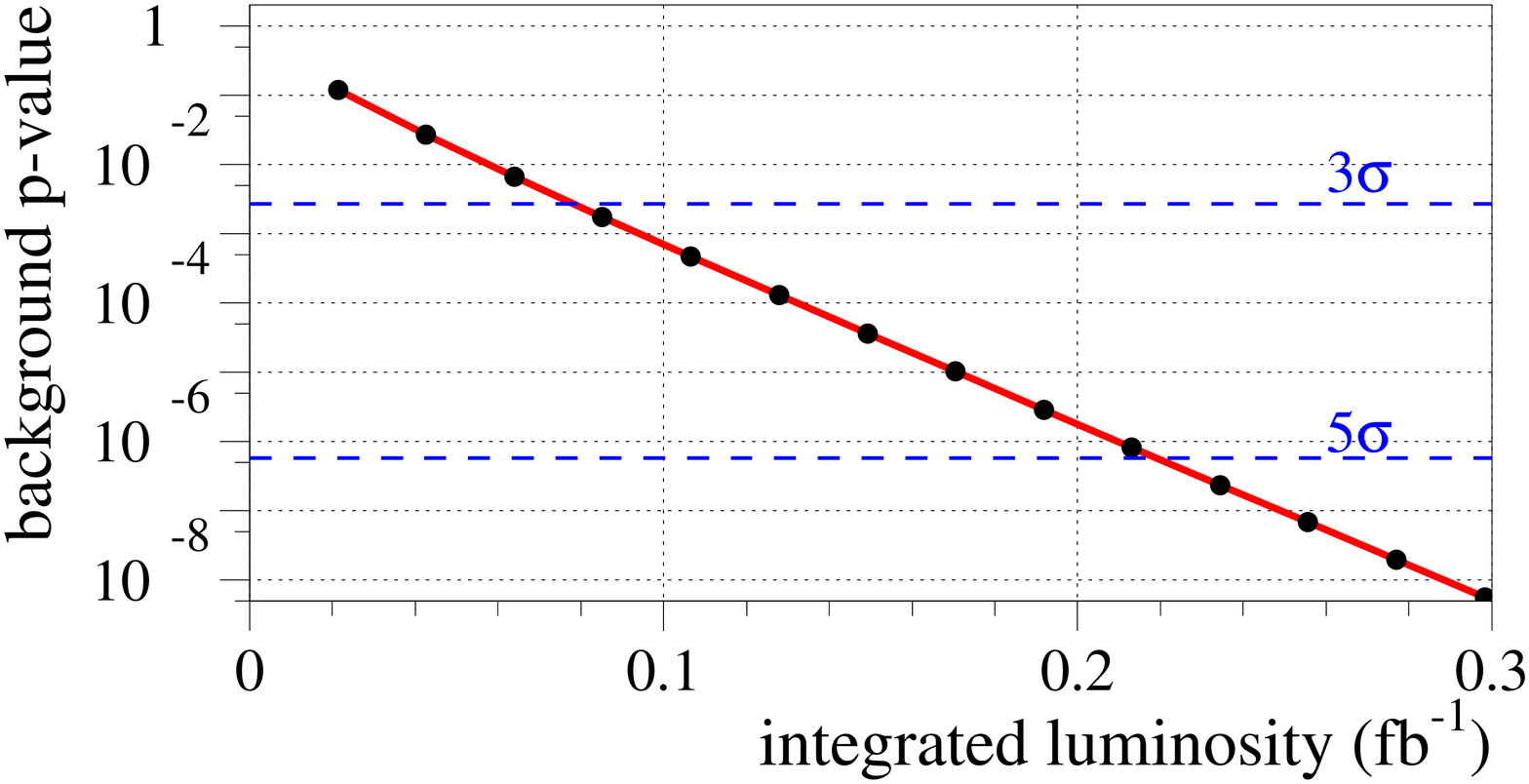}
\caption{\label{f:pvalues} $p$-values as a function of
integrated luminosity~$\Lum$. The black dots on the line 
show hypothetical integral numbers of observed events,
starting at $N = 1$ for $\Lum = 0.02~\fbinv$.}
}


\section{Summary}
\label{sec:concl}
A new method for a precise measurement of the stop quark mass
has been described, based on the ratio of yields at the peak 
stop quark pair production cross section, and near threshold.
This ratio is far less sensitive to experimental uncertainties than
other methods, leading to a very low estimated uncertainty,
still dominated by the statistical uncertainty and the theoretical
uncertainty (which is also present for other methods based on a
cross-section measurement).  We studied a specific scenario in
detail, with an emphasis on analysis techniques and
systematic uncertainties.  We placed special emphasis on the modeling
of the stop quark and charm fragmentation uncertainties, and suggest
how fragmentation models could be constrained with data taken at
the ILC.  Previous studies had not considered this source of uncertainty.
This method is general, and could be applied to other species,
provided an accurate prediction for the excitation curve is or
can be available. For weakly interacting particles, such as staus, 
the theoretical uncertainty is much smaller and the advantage of
the new method would be even more impressive.
\par
An important part of our studies is the use of multi-variate methods
to isolate a very clean stop quark signal.  For this we utilize the
Iterative Discriminant Analysis (IDA) used previously at LEP.
It is interesting that a carefully-tuned set of sequential cuts
achieves a much smaller systematic uncertainty, and hence a better overall
result for the stop quark mass measurement in this method.  The
superior background rejection of the IDA, however, is extremely useful
when searching for a stop signal, and we give an illustration for
$\sqrt{s} = 500\gev$, which shows that a five-sigma significance
could be obtained by the IDA selection with only $240~\pbinv$.
\par
The reduction of the uncertainty on the stop mass from about 
$\Delta\mst = 1.2\gev$ in Ref.~\cite{stop,heavyq} to
$\Delta\mst = 0.42\gev$ in this analysis is crucial for
testing theoretical explanations of the dark matter relic
density in the light-stop co-annihilation scenario. 
With these new results, the theoretical calculation has an accuracy
equal to the two-sigma uncertainty of the WMAP measurements.
The remaining uncertainty is no longer dominated by $\Delta\mst$.

\acknowledgments{
The authors are grateful to A.~Kraan, T.~Kuhl and S.~Mrenna for useful discussions and
expert advice on {\sc Pythia} and {\sc Simdet}.
 
ANL is supported by the U.S. Department of Energy (DoE) under Contract
DE-AC02-06CH11357. Fermilab is operated by Fermi Research Alliance, LLC under
Contract DE-AC02-07CH11359 with the U.S. Department of Energy.
The research effort by M.S. is supported by DoE Contract DE-FG02-91ER40684.
}


\end{document}